\documentclass{article}

\usepackage{arxiv}

\usepackage[utf8]{inputenc} 
\usepackage[T1]{fontenc}    
\usepackage{hyperref}       
\usepackage{url}            
\usepackage{booktabs}       
\usepackage{amsfonts}       
\usepackage{nicefrac}       
\usepackage{microtype}      
\usepackage{graphicx}
\usepackage{natbib}
\usepackage{doi}

\usepackage{hyperref}
\usepackage{amssymb}
\usepackage{amsmath}
\usepackage{subcaption}
\usepackage[frozencache=true,cachedir=minted-cache]{minted}
\usepackage{dirtytalk}

\hypersetup{
    colorlinks=true,
    linkcolor=blue,
    urlcolor=cyan,
    citecolor=black,
    }

\newcommand{\nn}{N\!N}

\title{Finite Basis Physics-Informed Neural Networks (FBPINNs): a scalable domain decomposition approach for solving differential equations}

\author{ 
    Ben Moseley\thanks{bmoseley@robots.ox.ac.uk} \\
	Department of Computer Science\\
	University of Oxford\\
	Oxford, UK\\
	\And
    Andrew Markham \\
	Department of Computer Science\\
	University of Oxford\\
	Oxford, UK\\
	\And
    Tarje Nissen-Meyer \\
	Department of Earth Sciences\\
	University of Oxford\\
	Oxford, UK\\
}

\renewcommand{\shorttitle}{FBPINNs: a scalable domain decomposition approach for solving differential equations}

\hypersetup{
pdftitle={Finite Basis Physics-Informed Neural Networks (FBPINNs): a scalable domain decomposition approach for solving differential equations},
}

\begin{document}
\maketitle

\begin{abstract}
Recently, physics-informed neural networks (PINNs) have offered a powerful new paradigm for  solving problems relating to differential equations. Compared to classical numerical methods PINNs have several advantages, for example their ability to provide mesh-free solutions of differential equations and their ability to carry out forward and inverse modelling within the same optimisation problem. Whilst promising, a key limitation to date is that PINNs have struggled to accurately and efficiently solve problems with large domains and/or multi-scale solutions, which is crucial for their real-world application. Multiple significant and related factors contribute to this issue, including the increasing complexity of the underlying PINN optimisation problem as the problem size grows and the spectral bias of neural networks. In this work we propose a new, scalable approach for solving large problems relating to differential equations called \emph{Finite Basis PINNs (FBPINNs)}. FBPINNs are inspired by classical finite element methods, where the solution of the differential equation is expressed as the sum of a finite set of basis functions with compact support. In FBPINNs neural networks are used to learn these basis functions, which are defined over small, overlapping subdomains. FBINNs are designed to address the spectral bias of neural networks by using separate input normalisation over each subdomain, and reduce the complexity of the underlying optimisation problem by using many smaller neural networks in a parallel divide-and-conquer approach. Our numerical experiments show that FBPINNs are effective in solving both small and larger, multi-scale problems, outperforming standard PINNs in both accuracy and computational resources required, potentially paving the way to the application of PINNs on large, real-world problems.
\end{abstract}

\keywords{Physics informed neural networks \and Domain decomposition \and Learnable basis functions \and Spectral bias issue \and Multi-scale modelling \and Forward modelling \and Differential equations  \and Parallel computing}



\section{Introduction}
\label{sec:introduction}

Solving forward and inverse problems relating to differential equations has been the subject of intense research focus over many decades. Indeed, many areas of science rely on our ability to accurately solve these problems, from modelling the climate to understanding fundamental particle physics \citep{Giorgi2019, Prein2015, Agostinelli2003}. Classical approaches such as finite difference, finite element, and spectral methods have emerged as the dominant approach, with great advances in their performance being made throughout the decades. Today these methods are capable of solving highly complex, large-scale problems, and are often implemented using sophisticated, adaptive and parallel algorithms utilising thousands of CPU and GPU cores \citep{Jasak2009a, Nissen-Meyer2014}.

Whilst powerful, classical methods also have some long-standing limitations. For example, many of today's applications require the study of multi-physics, multi-scale systems, which can be very challenging to incorporate and model accurately, often requiring the use of adaptive schemes and subgrid parameterisations \citep{Nochetto2009, Rasp2018}. Another is the vast computational resources that are typically required, which render many real-world applications unfeasible. Thirdly, implementations of classical approaches are often elaborate and can require millions of lines of code, making them challenging to maintain from generation to generation \citep{Jasak2009a}.

In recent years researchers have turned towards the rapid advances in machine learning as a potential way to address some of these challenges. One strategy is to entirely replace classical methods with purely data-driven algorithms (for example see \cite{Moseley2020a, Kasim2020}) and whilst this leads to advantages such as dramatic savings in computational efficiency, it reveals important flaws in current machine learning techniques such as issues with generalisation and the need for large amounts of training data. A more recent strategy is to blend together our physical principles and traditional algorithms with machine learning in a more nuanced way to create more powerful models, in the burgeoning field of scientific machine learning (SciML) \citep{Baker2019}.

One such approach which has received significant attention are \emph{physics-informed neural networks} (PINNs) \citep{Lagaris1998, Raissi2019, Karniadakis2021}, which can be used to solve both forward and inverse problems relating to differential equations. PINNs use a deep neural network to represent the solution of a differential equation, which is trained using a loss function which directly penalises the residual of the underlying equation. PINNs have multiple benefits in comparison to classical methods, for example they provide approximate mesh-free solutions which have tractable analytical gradients, and they provide an elegant way to carry out joint forward and inverse modelling within the same optimisation problem \citep{Raissi2019}. Many of the fundamental concepts behind PINNs were proposed in the 1990s by \cite{Lagaris1998} and others, whilst \cite{Raissi2019} implemented and extended them with modern deep learning techniques.

Since these works, PINNs have been utilised in a wide range of applications, for example in simulating fluid flow \citep{Raissi2020, Sun2019a}, carrying out cardiac activation mapping \citep{SahliCostabal2020}, modelling wave physics \citep{Moseley2020c}, and even in predicting the temperature distribution over an espresso cup \citep{Cai2021}. Many extensions of PINNs have also been investigated. For example, one extension is to include uncertainty estimation within PINNs, where  \cite{Yang2021} proposed Bayesian PINNs which use Bayesian neural networks to estimate uncertainty in the PINN solution. Another is to use the variational form of differential equations to construct an alternative optimisation problem (for example, see \cite{Kharazmi2019}). \cite{Zhu2019b}, \cite{Geneva2020a} and \cite{Gao2021} instead used finite difference filters to construct the differential equation residual in a PINN-like loss function when training CNNs and RNNs  which approximate the solution of a differential equation on a mesh. The advantage of this approach is that it allows these models to be easily conditioned on the initial conditions of the problem. Similarly, \cite{Lu2021a} proposed DeepONets, which condition PINNs on their initial conditions such that they learn families of solutions. PINNs have been extended to fractional differential equations \citep{Pang2019}, stochastic differential equations \citep{Yang2018, Zhang2019b}, and have even been used to discover the underlying differential equations themselves \citep{Chen2020c}. Multiple software libraries have been developed which allow PINNs to be trained easily and quickly, such as DeepXDE \citep{Lu2019}, SimNet \citep{Hennigh2020}, PyDEns \citep{Koryagin2019} and NeuroDiffEq \citep{Chen2020a}.

Whilst popular and effective, PINNs also suffer from some significant limitations. One is that, in comparison to classical approaches, the theoretical convergence properties of PINNs are still poorly understood. For example, the PINN loss function can be highly non-convex and result in a stiff optimisation problem, yet it is unclear which classes of problems this affects \citep{Wang2020, Wang2020d}. Work by \cite{Wang2020d, Shin2020, Mishra2020a} and others have made initial steps towards understanding the theoretical properties of PINNs but this sub-field is in its early stages. Another is the poor computational efficiency of PINNs; a standard PINN must be retrained for each solution, which is expensive and typically means classical methods strongly outperform PINNs, at least for forward modelling tasks. Works which train PINNs to learn a family of solutions, such as the aforementioned DeepONets, could potentially address this issue.

Finally, a major challenge is scaling PINNs to large problem domains. There are multiple related issues in this respect. One is that, as the domain size increases, the complexity of the solution typically increases, which inevitably requires the size of the neural network (or number of free parameters) to increase such that the network is expressive enough to represent it. This in turn results in a harder PINN optimisation problem, both in terms of the number of free parameters and the increased number of training points required to sufficiently sample the solution over the larger domain, and typically leads to much slower convergence times. Another is the \emph{spectral bias} of neural networks. This is the well-studied observation that neural networks tend to learn higher frequencies much more slowly than lower frequencies \citep{Xu2019, Rahaman2018, Basri2019, Cao2019}, with various convergence rates being proved. Because a problem's input variables are typically normalised over the domain before being input to a PINN, low frequency features in the solution effectively become high frequency features as the domain size increases, which can severely hinder the convergence of PINNs. Similarly, spectral bias is a major issue when solving problems with multi-scale solutions \citep{Wang2020e}. These limitations have meant that the vast majority of PINN applications to date have only solved problems in small domains with limited frequency content.

Recent works have started to investigate these issues. \cite{Wang2020e} showed that spectral bias also affects PINNs, and proposed the use of Fourier input features to help alleviate this issue, which transform the input variables over multiple scales using trigonometric functions as a pre-processing step. \cite{Liu2020} proposed multi-scale deep neural networks, which used radial scaling of the input variables in the frequency domain to achieve uniform convergence across multiple scales when solving problems with a PINN loss. However, for both these methods, the scale of their input features must be chosen to match the frequency range of interest. 

For reducing the complexity of the PINN optimisation problem, an increasingly popular approach is to use \emph{domain decomposition} \citep{Heinlein2021}, taking inspiration from existing classical methods such as finite element modelling. Instead of solving one large PINN optimisation problem, the idea is to use a \say{divide and conquer} approach and train many smaller problems in parallel. This can reduce training times and could potentially reduce the difficulty of the global optimisation problem too. For example, \cite{Jagtap2020b} proposed XPINNs, which divide the problem domain into many subdomains, and use separate neural networks in each subdomain to learn the solution. A key consideration in any domain decomposition approach is to ensure that individual subdomain solutions are communicated and match across the subdomain interfaces, and \cite{Jagtap2020b} rely upon additional interfaces terms in their PINN loss function to do so. In follow up work \cite{Shukla2021} showed their approach can be parallelised to multiple GPUs, decreasing training times. However, a key downside of their approach is that it contains discontinuities in the PINN solution across subdomain interfaces, as the interface conditions are only weakly constrained in the loss function. Similar domain decomposition strategies were proposed by \cite{Dwivedi2021} and \cite{Dong2020}, but who instead used extreme learning machines (ELMs) \citep{Huang2006} as the neural networks in each sudomain, which can be rapidly trained. However, whilst \cite{Dong2020} claimed computational efficiency comparable to finite element modelling, ELMs restrict the capacity of neural networks as only the weights in the last layer can be updated. In other related approaches, \cite{Li2020e} replaced subdomain solvers of the classical Schwarz domain decomposition approach with PINNs, resulting in a similar strategy to \cite{Jagtap2020b}. \cite{Stiller2020} proposed GatedPINNs, which use an auxiliary neural network to learn the domain decomposition itself, and \cite{Kharazmi2021} use domain decomposition to train variational PINNs, although only when defining the space of test functions.

In this work, we propose a new domain decomposition approach for solving large, multi-scale problems relating to differential equations called \emph{Finite Basis PINNs (FBPINNs)}. In contrast to the existing domain decomposition approaches, we ensure that interface continuity is strictly enforced across the subdomain boundaries by mathematical construction of our PINN solution ansatz, which removes the need for additional interface terms in our PINN loss function. Furthermore, we explicitly consider the effects of spectral bias by using separate input variable normalisation within each subdomain, and we aid the convergence of FBPINNs by using flexible training schedules. We further propose a parallel training algorithm for FBPINNs allowing them to be scaled computationally. To numerically validate our approach, we compare the accuracy of FBPINNs to standard PINNs across a range of problem sizes, both in terms of domain size and dimensionality, including problems with multi-scale solutions. These experiments suggest that FBPINNs are effective in solving both small and larger, multi-scale problems, outperforming standard PINNs in both accuracy and computation resources required. All of the code for reproducing the results of this paper can be found at: \href{https://github.com/benmoseley/FBPINNs}{github.com/benmoseley/FBPINNs}.

The remainder of this paper is as follows. In Section~\ref{sec:PINN} we describe the standard PINN implementation as defined by \cite{Raissi2019}, as well as the alternative constructed ansatz formulation proposed by \cite{Lagaris1998}. In Section~\ref{sec:motivation} we motivate FBPINNs by showing a simple example where PINNs break down; that of learning the solution to $\frac{du}{dx} = \cos \omega x$ for large values of $\omega$. In Section~\ref{sec:methods} we present the FBPINN methodology in detail. In Section~\ref{sec:results} we present numerical results which compare the performance PINNs and FBPINNs across many smaller and larger scale problems, including the Burgers equation and the wave equation. Finally in Sections~\ref{sec:discussion} and \ref{sec:conclusion} we discuss the implications and conclusions of our work.


\section{Physics informed neural networks (PINNs)}
\label{sec:PINN}

PINNs are designed to solve differential equations of the general form \citep{Raissi2019}
\begin{equation}
\begin{split}
\mathcal{D}[u(x); \lambda] &= f(x)~,~~~~x \in \Omega~,\\
\mathcal{B}_{k}[u(x)] &= g_{k}(x)~,~~~x \in \Gamma_{k} \subset \partial \Omega~,
\label{eq:problem}
\end{split}
\end{equation}
for $k=1,2,...,n_{b}$ where $\mathcal{D}$ is a differential operator, $\mathcal{B}_{k}$ is a set of boundary operators, $u \in \mathbb{R}^{d_{u}}$ is the solution to the differential equation, $f(x)$ is a forcing function, $g_{k}(x)$ is a set of boundary functions, $x$ is an input vector in the domain $\Omega \subset \mathbb{R}^{d}$ (i.e. $x$ is a $d$-dimensional vector), $\partial \Omega$ denotes the boundary of $\Omega$ and $\lambda$ is an optional vector of additional parameters of the differential operator. Many different physical systems can be described in this form, including problems with time-dependence or time-independence, linear or nonlinear differential operators, and different types of initial and boundary conditions such as Dirichlet and Neumann conditions. As an example, the inhomogeneous wave equation with Dirichlet and Neumann boundary conditions at $t=0$ reads
\begin{equation}
\begin{split}
\left[ \nabla^{2} - \frac{1}{c^2}\frac{\partial^2}{\partial t^2} \right] u(x,t) &= f(x,t)~,\\
u(x,0) &= g_{1}(x)~,\\
\frac{\partial u}{\partial t} (x,0) &= g_{2}(x)~,
\label{eq:problem_wave_intro}
\end{split}
\end{equation}
where $c$ is the wave speed (note, for all time-dependent problems described in this paper we have labelled the time dimension separately from the spatial dimensions of the input vector for readability).

Given the problem definition in Equation~\ref{eq:problem}, PINNs train a neural network, denoted as $\nn(x; \theta)$ where $\theta$ denotes the free parameters of the neural network, to directly approximate the solution to the differential equation, ie $\nn(x;\theta) \approx u(x)$. To solve the problem the network is trained by minimising the following loss function;
\begin{equation}
\mathcal{L}(\theta) = \mathcal{L}_{p}(\theta) + \mathcal{L}_{b}(\theta)~,
\label{eq:pinn}
\end{equation}
where
\begin{align}
\mathcal{L}_{p}(\theta) &= \frac{1}{N_p}\sum^{N_p}_{i} ||\,\mathcal{D}[\nn(x_i;\theta); \lambda] - f(x_i)\,||^2~, \label{eq:physics} \\
\mathcal{L}_{b}(\theta) &= \sum_{k} \frac{1}{N_{bk}}\sum^{N_{bk}}_{j} ||\,\mathcal{B}_{k}[\nn(x_{kj};\theta)] - g_{k}(x_{kj})\,||^2~, \label{eq:boundary}
\end{align}
$\{x_i\}$ is a set of training points sampled over the full domain $\Omega$ and $\{x_{kj}\}$ is a set of training points sampled from the boundary $\Gamma_{k}$ associated with each boundary condition. We denote the first term, $\mathcal{L}_{p}(\theta)$, as the \say{physics loss} and the second term, $\mathcal{L}_{b}(\theta)$, as the \say{boundary loss}. Intuitively, the physics loss pushes the neural network to learn solutions which are consistent with the underlying differential equation, whilst the boundary loss attempts to ensure the solution is unique by matching the solution to the boundary conditions. For the wave equation example above, Equation~\ref{eq:pinn} becomes
\begin{equation}
\begin{split}
& \mathcal{L}(\theta) = \frac{1}{N_p}\sum^{N_p}_{i} ||\,\left[ \nabla^{2} - \frac{1}{c^2}\frac{\partial^2}{\partial t^2} \right]\nn(x_i,t_i;\theta) - f(x_i,t_i)\,||^2 \\
& + \frac{1}{N_{b1}}\sum^{N_{b1}}_{j} ||\,\nn(x_{1j},0;\theta) - g_1(x_{1j})\,||^2 + \frac{1}{N_{b2}}\sum^{N_{b2}}_{j} ||\,\frac{\partial}{\partial t}\nn(x_{2j},0;\theta) - g_2(x_{2j})\,||^2~.
\label{eq:pinn_wave_intro}
\end{split}
\end{equation}

There are a number of important points to consider when training PINNs. The first is that a sufficient number of training points $\{x_i\}$ and $\{x_{kj}\}$ should be selected such that the network is able to learn a consistent solution across the entire domain. For large domains, training points are often sampled in mini-batches and the optimisation problem becomes a stochastic optimisation problem. Secondly, when evaluating the loss function the gradients of the neural network with respect to its inputs are required. These are typically analytically available and this allows the loss function to be evaluated and further differentiated with respect to $\theta$, allowing gradient descent methods to be used to optimise $\theta$. In practice these gradients are easily obtainable in modern deep learning packages through the use of autodifferentiation \citep{pytorch}. Thirdly, it is important to note that whilst the known values of the solution (and/or its derivatives) are required at the domain boundaries to evaluate the boundary loss, evaluating the physics loss only requires samples of the input vector, i.e. it can be viewed as an unsupervised regularisation term, and therefore PINNs require very little training data in this sense. Finally, PINNs can be naturally extended to inverse problems, for example when estimating the parameters of the differential equation $\lambda$. In this setting, an additional \say{data loss} is usually added, which penalises the difference between the network solution and the solution at a set of observed points within the domain, and the parameters $\lambda$ are jointly optimised alongside $\theta$. 

\subsection{Weakly vs strongly constrained PINNs}
\label{sec:hard_pinns}

A downside of the PINN approach described above is that the boundary conditions are only weakly enforced in the loss function described by Equation~\ref{eq:pinn}, meaning that the learned solution may be inconsistent. Furthermore recent work has shown theoretically and empirically that the PINN optimisation problem can be stiff to solve, to the point where it may not converge at all, due to the two terms in the loss function competing with each other \citep{Wang2020, Wang2020d, Sun2019a}. An alternative approach, as originally proposed by \cite{Lagaris1998}, is to strictly enforce the boundary conditions by using the neural network as part of a solution ansatz. For example, for the wave equation problem above, instead of defining a neural network to directly approximate the solution $\nn(x,t;\theta) \approx u(x,t)$, one could instead use the ansatz
\begin{equation}
\hat u(x, t; \theta) = g_1(x) + t \, g_2(x) + t^2 \, \nn(x,t;\theta)~,
\label{eq:ansatz_wave_intro}
\end{equation}
to define the approximate solution in the PINN optimisation problem. It is straightforward to verify that this ansatz automatically satisfies the boundary conditions in Equation~\ref{eq:problem_wave_intro}. Because the boundary conditions are automatically satisfied, only the physics loss, $\mathcal{L}_{p}(\theta)$, needs to be included in Equation~\ref{eq:pinn}, which turns the optimisation problem from a constrained one into a simpler unconstrained one. This formulation has since been extended to irregular domains \citep{Berg2018}, and general schemes for constructing suitable ansatze have been proposed \citep{Leake2020}. We will use this strategy for the rest of the examples in the paper, although we note that the FBPINN framework is not limited to this approach and can use both formulations.

\section{A motivating example}
\label{sec:motivation}

\begin{figure}[t]
\begin{center}
\includegraphics[width=15cm]{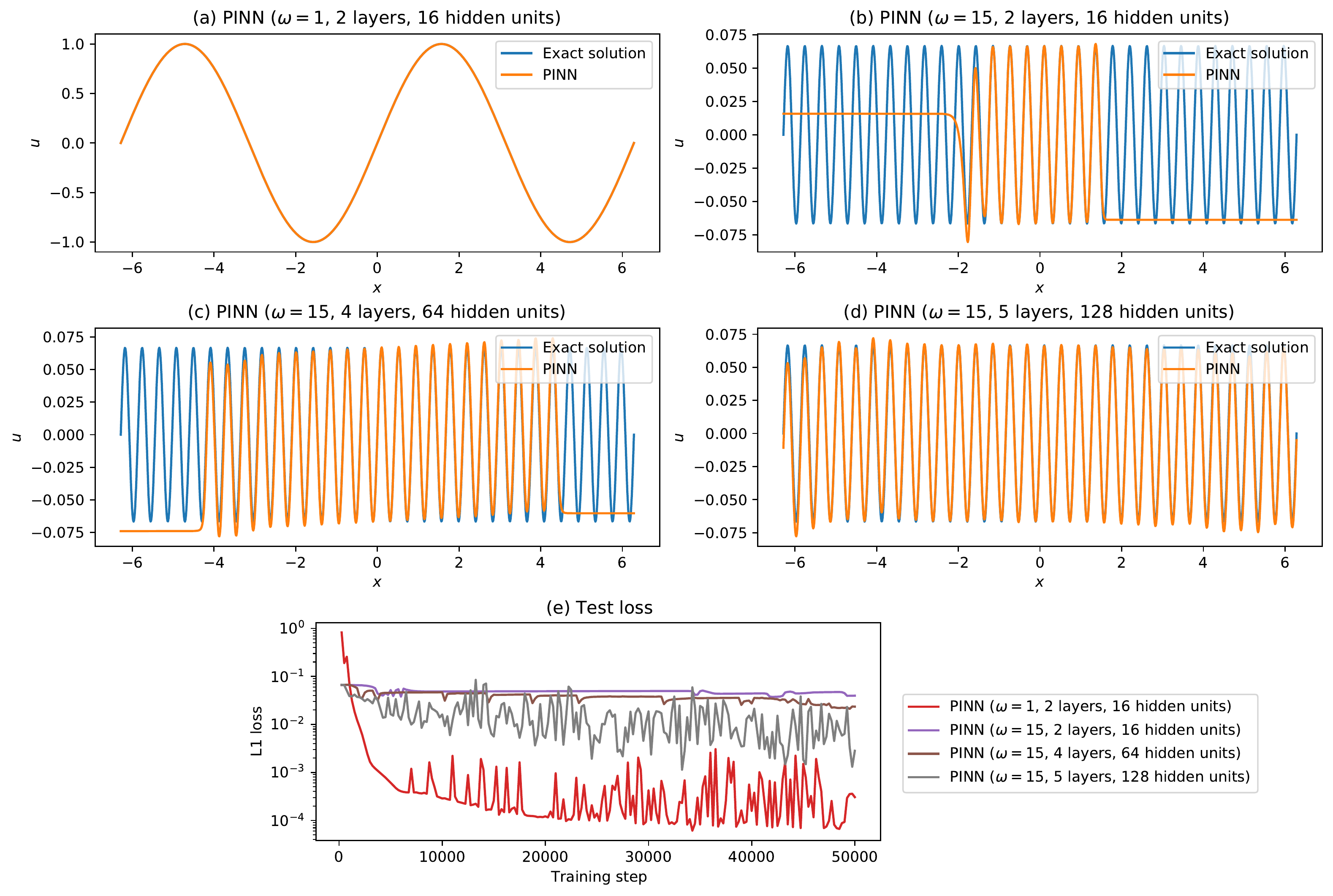}
\caption[]{A motivating problem: using PINNs to solve $\frac{d u}{d x} = \cos(\omega x)$. When $\omega=1$ (i.e. $\omega$ is low), a PINN with 2 hidden layers and 16 hidden units is able to rapidly converge to the solution; (a) shows the PINN solution compared to the exact solution and (e) shows the L1 error between the PINN solution and the exact solution against training step. When $\omega=15$ (i.e. $\omega$ is high), the same PINN struggles to converge, as shown in (b). Whilst increasing the size (number of free parameters) of the PINN improves its accuracy, as shown in (c) and (d), it converges much more slowly and with much lower accuracy than the low frequency case, as shown in (e).}
\label{fig:cosp}
\end{center}
\end{figure}

As discussed in Section~\ref{sec:introduction}, a major challenge is to scale PINNs to large domains. In this section, we will show a simple empirical 1D example highlighting this difficulty, which motivates the FBPINN framework. Specifically, we consider the following problem
\begin{equation}
\begin{split}
\frac{du}{dx} &= \cos(\omega x)~,\\
u(0) &= 0~,
\end{split}
\label{eq:problem_motivate}
\end{equation}
where $x,u,\omega \in \mathbb{R}^{1}$, which has the exact solution
\begin{equation}
u(x) = \frac{1}{\omega} \sin(\omega x)~.
\end{equation}
We will use the PINN solution ansatz
\begin{equation}
\hat u(x; \theta) = \tanh(\omega x) \nn(x;\theta)~,
\label{eq:ansatz_motivate}
\end{equation}
to solve this problem. We choose to use the $\tanh$ function in the ansatz because away from the boundary condition its value tends to $\pm 1$, such that the neural network does not need to learn to dramatically compensate for this function away from the boundary. We also approximately match the width of the $\tanh$ function to the wavelength of the exact solution, again such that the compensation the network needs to learn in the vicinity of the boundary is of similar frequency to the solution. In our subsequent experiments we find both of these strategies help PINNs converge over large domains and different solution frequencies.

The PINN is trained using the unconstrained loss function
\begin{equation}
\mathcal{L}(\theta) = \mathcal{L}_{p}(\theta) = \frac{1}{N_p}\sum^{N_p}_{i} ||\, \frac{d}{dx} \hat u (x_i;\theta) - \cos(\omega x_i)\,||^2~,
\end{equation}
as derived using Equation~\ref{eq:physics}.

\subsection{Low frequency case ($\omega=1$)}
\label{sec:motivation_low}

First, the above PINN is trained for the $\omega$ = 1 case. We use a fully connected network with 2 hidden layers, 16 hidden units per layer and $\tanh$ activation functions, using 200 training points ($\{x_i\}$) regularly spaced over the domain $x\in [-2 \pi,2 \pi]$. The input variable $x$ is normalised to $[-1,1]$ over the domain before being input to the network, and the output of the network is unnormalised by multiplying it by $\frac{1}{\omega}$ before being used in the ansatz defined by Equation~\ref{eq:ansatz_motivate}. The PINN is trained using gradient descent with the Adam optimiser \citep{Kingma2014} and a learning rate of 0.001. All code is implemented using the \verb+PyTorch+ library, using its autodifferentiation features \citep{pytorch}.

Figure~\ref{fig:cosp}~(a) and (e) shows the PINN solution after training and its L1 error compared to the exact solution (evaluated using 1000 regularly spaced points in the domain) as a function of training step. We find that the PINN is able to quickly and accurately converge to the exact solution.

\subsection{High frequency case ($\omega=15$)}
\label{sec:motivation_high}

Next, the same experiment is run, but with two changes; we increase the frequency of the solution to $\omega=15$ and the number of regularly spaced training points to $200 \times 15 = 3000$. Figure~\ref{fig:cosp}~(b) and (e) shows the resulting PINN solution and L1 convergence curve (using 5000 test points). We find for this case the PINN is unable to accurately learn the solution, only being able to capture the first few cycles of the solution away from the boundary condition. We further retrain the PINN using larger network sizes of 4 layers and 64 hidden units, and 5 layers and 128 hidden units, shown in Figure~\ref{fig:cosp}~(c) and (d), and find that only the PINN with 5 layers and 128 hidden units is able to fully model all of the cycles, although its final relative L1 error is much worse than the $\omega=1$ case and its convergence curve is much slower and more unstable. We note that the $\omega = 1$ PINN with 2 layers and 16 hidden units uses 321 free parameters whilst the $\omega = 15$ PINN with 5 layers and 128 hidden units uses 66,433 free parameters.

\subsection{Remarks}

Whilst the PINN performs well for low frequencies, it struggles to scale to higher frequencies; the higher frequency PINN requires many more free parameters, converges much more slowly and has worse accuracy. Multiple significant and related factors contribute to this issue. One is that as the complexity of the solution increases, the network requires more free parameters to accurately represent it, making the optimisation problem harder. Another is that as the frequency increases, more training sample points are required to sufficiently sample the domain, again making the optimisation problem harder. Third is the \emph{spectral bias} of neural networks, which is the observation that neural networks tend to learn higher frequencies much more slowly than low frequencies, a property which has been rigorously studied. Compounding these factors is the fact that, as the size of the network, number of training points, and convergence time grows, the computational resources required increases significantly.

It is important to note that for this problem, scaling to higher frequencies is equivalent to scaling to larger domain sizes. Because we have normalised the input variable to $[-1,1]$ within the domain before inputting it to the network, keeping $\omega = 1$ but expanding the domain size by 15 times and re-normalising presents the same optimisation problem to the neural network as changing $\omega$ to $15$. Indeed, increasing the domain size and scaling to higher frequencies are related problems, and the above case study highlights a general observation of PINNs: as the domain size increases, the PINN optimisation typically becomes much harder and takes much longer to converge. 

Another important note is that classical methods also typically scale poorly to large domain sizes/ higher frequencies. For example, when solving the problem above the number of mesh points required by standard finite difference modelling would scale $\propto \omega^{d}$. However, FD modelling would not suffer from the additional PINN-related problems described above of many more free parameters in the optimisation problem and slower convergence due to spectral bias. 

In the next section we will present FBPINNs, which, as we shall see in Section~\ref{sec:results}, are able to solve the case study above much more accurately and efficiently than the PINNs studied.


\section{Finite Basis PINNs (FBPINNs)}
\label{sec:methods}

\begin{figure}[t]
\begin{center}
\includegraphics[width=15cm]{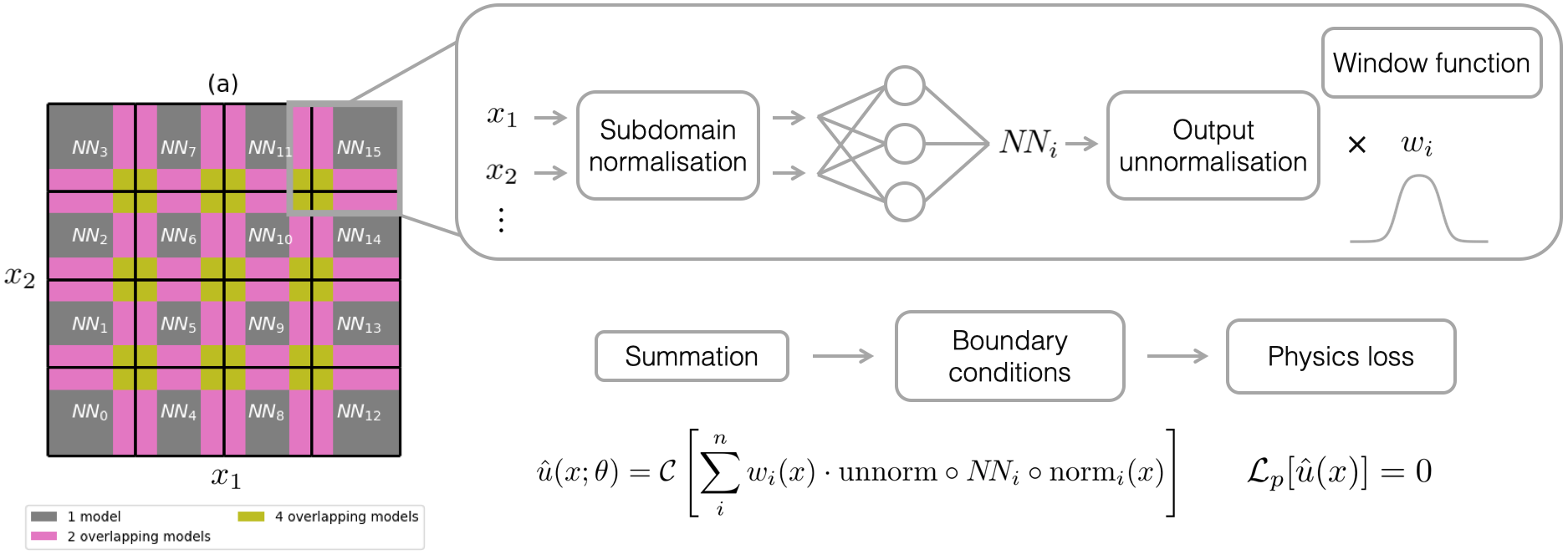}
\caption[]{FBPINN workflow. FBPINNs use domain decomposition and separate subdomain normalisation to address the issues related to scaling PINNs to large domains. First, the problem domain is divided into overlapping subdomains; an example 2D hyperrectangular subdivision is shown in (a). Next, separate neural networks are placed within each subdomain. Each network is locally confined to its subdomain by multiplying it with a differentiable window function, such that within the center of the subdomain, the network learns the full solution to the differential equation, whilst in the overlapping regions, the solution is defined as the sum over all overlapping networks. For each network, the input variables $x = (x_1$, $x_2, ...)$ are normalised between [-1,1] over the subdomain, and the output of the network is unnormalised using a common unnormalisation. Finally, an optional constraining operator $\mathcal{C}$ can be applied which appropriately constrains the ansatz such that it automatically satisfies the boundary conditions. FBPINNs are trained using a very similar loss function to standard PINNs which does not require the use of additional interface terms.}
\label{fig:workflow}
\end{center}
\end{figure}

\begin{figure}[t]
\begin{center}
\includegraphics[width=13cm]{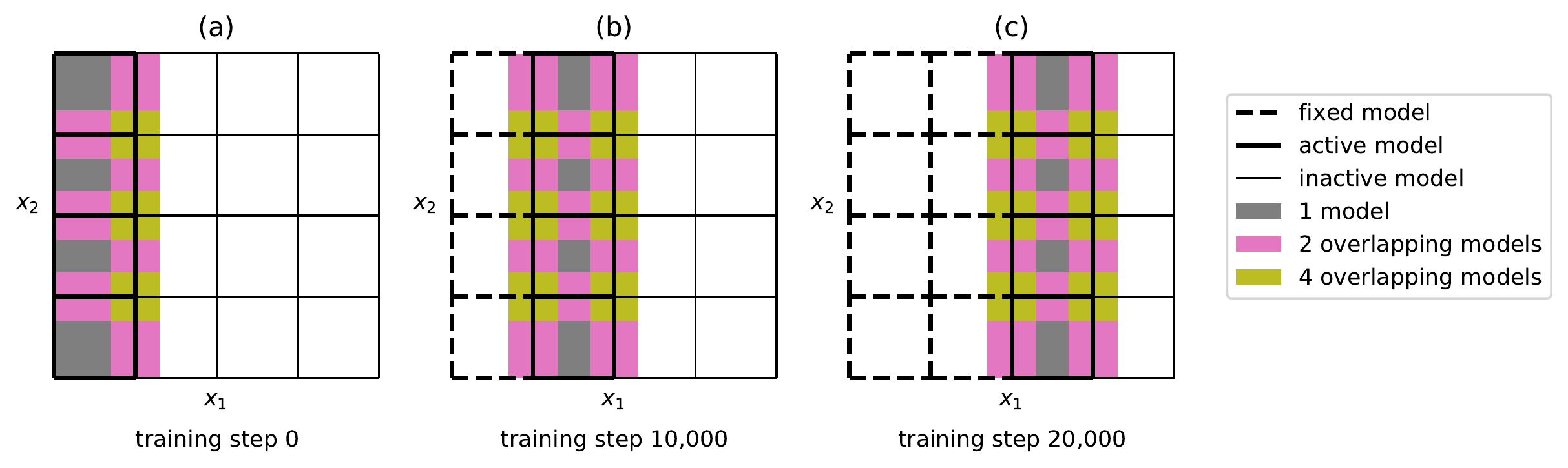}
\caption[]{Flexible training schedules for FBPINNs. We can design flexible training schedules which can help to improve the convergence of FBPINNs. These schedules define which subdomain networks are updated during each training step. Within these schedules we define \say{active} models, which are the networks which are currently being updated, \say{fixed} models, which are networks which have already been trained and have their free parameters fixed, and \say{inactive} models which are as-of-yet untrained networks. The plots above show one particular training schedule designed to learn the solution \say{outwards} from the boundary condition, which in this case is assumed to be along the left edge of the domain. Note during each training step only training points from the active subdomains are required, shown by the coloured regions in the plot.}
\label{fig:active}
\end{center}
\end{figure}

\begin{figure}
\centering
\SaveVerb{pytorch}|PyTorch|
\begin{subfigure}{8.6cm}
\centering
\begin{minted}[
frame=lines,
framesep=2mm,
fontsize=\footnotesize,
]{python}
# 1) Get NN output and gradients in each subdomain
# Can use separate thread for each subdomain
for im in active_or_fixed_neighbour_models:
    x[im] = sample_subdomain_points(im)
    x_norm = (x[im] - x_mu[im]) / x_sd[im]
    u[im] = NN[im](x_norm)
    u[im] = u[im]*u_mu + u_sd
    u[im] = windows[im] * u[im]
    g[im] = problem.get_gradients(u[im], x[im])

# 2) Sum NN outputs in overlapping regions
# Requires communication between threads
for im in active_models:
    for iseg in overlapping_regions[im]:
        for im2 in overlapping_models[iseg]:
            if im2 != im:
                u[im][iseg] += u[im2][iseg].detach()
                g[im][iseg] += g[im2][iseg].detach()

# 3) Apply hard boundary conditions, compute loss,
# backpropagate and update free parameters
# Can use separate thread for each subdomain
for im in active_models:
    u[im], g[im] = problem.boundary(u[im], g[im])
    loss = problem.physics_loss(x[im], u[im], g[im])
    loss.backward()
    optimizers[im].step()
\end{minted}
\caption{\protect\UseVerb{pytorch} psuedocode for a single FBPINN training step.}
\end{subfigure}%
~~
\begin{subfigure}{8cm}
\centering
\includegraphics[width=\linewidth]{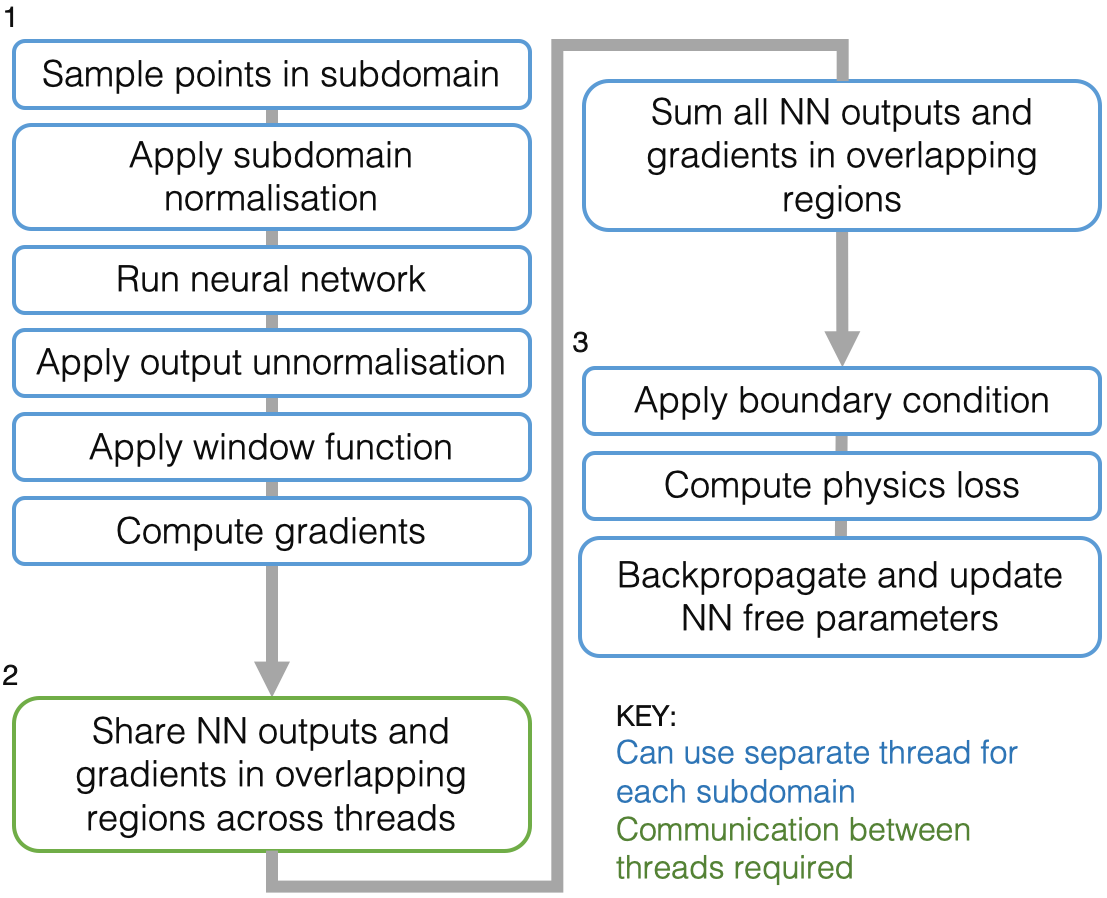}
\caption{Visual schematic of a FBPINN training step inside each subdomain.}
\end{subfigure}
\caption{Parallel algorithm for training FBPINNs. FBPINNs are trained using gradient descent, and the psuedocode for each training step is shown in (a). The effect of each training step on each subdomain is shown in (b). The algorithm can be implemented using entirely independent threads for each subdomain, except during step 2) where they must share their subdomain network outputs within overlapping regions with the threads of their neighbouring subdomains.}
\label{fig:threads}
\end{figure}

\subsection{Workflow overview}

FBPINNs are a general domain decomposition approach for solving large, multi-scale problems relating to differential equations. The main goal of FBPINNs is to address the scaling issues of PINNs described above, which is achieved by using a combination of domain decomposition, subdomain normalisation and flexible training schedules. In this subsection we give an overview of FBPINNs before giving a detailed mathematical description in Section~\ref{sec:methods_maths}.

The FBPINN workflow is shown in Figure~\ref{fig:workflow}. In FBPINNs, the problem domain is divided into overlapping subdomains. A neural network is placed within each subdomain such that within the center of the subdomain, the network learns the full solution, whilst in the overlapping regions, the solution is defined as the sum over all overlapping networks. Before being summed each network is multiplying by a smooth, differentiable window function which locally confines it to its subdomain. In addition to this the input variables of each network are separately normalised over each subdomain, and we can define flexible training schedules which allow us to restrict which subdomain networks are updated at each gradient descent training step (described in more detail in Section~\ref{sec:methods_schedules}).

By dividing the domain into many subdomains the single, large PINN optimisation problem is turned into many smaller subdomain optimisation problems. By using separate subdomain normalisation (as well as a domain decomposition appropriate for the complexity of the solution) we ensure that the effective solution frequency each subdomain optimisation problem sees is low. The use of flexible training schedules allow us to focus on solving smaller parts of the domain sequentially instead of all of them at once. Our hypothesis is that all three of these strategies alleviate the scaling issues described above, leading to an easier global optimisation problem.

With any domain decomposition technique it is important to ensure that the individual neural network solutions are communicated and match across the subdomain interfaces. In FBPINNs, during training the neural networks share their solutions in the overlap regions between subdomains, and by constructing our global solution as the sum of these solutions we automatically ensure the solution is continuous across subdomains. This approach allows FBPINNs to use a similar loss function to PINNs, without requiring the use of additional interface loss terms, in contrast to other domain decomposition approaches (e.g. \cite{Jagtap2020b}). 

Finally, FBPINNs can be trained in a highly parallel fashion, decreasing their training times, and we present an algorithm for doing so in Section~\ref{sec:parallel}. FBPINNs are inspired by classical finite element methods (FEMs), where the solution of the differential equation is expressed as the sum of a finite set of basis functions with compact support, although we note that FBPINNs use the strong form of the governing equation as opposed to the weak form in FEMs.

\subsection{Mathematical description}
\label{sec:methods_maths}

We will now give a detailed mathematical description of FBPINNs. In FBPINNs, the problem domain $\Omega \subset \mathbb{R}^{d}$ is subdivided into $n$ overlapping subdomains, $\Omega_{i} \subset \Omega$. FBPINNs can use any type of subdivision, regular or irregular, with any overlap width, as long as the subdomains overlap. An example subdivision, a regular hyperrectangular division, is shown in Figure~\ref{fig:workflow} (a). For simplicity, we use this type of division for the rest of this work.

Given a subdomain definition, the following FBPINN solution ansatz is used to define the approximate solution to the problem defined by Equation~\ref{eq:problem}
\begin{equation}
\hat u(x; \theta) = \mathcal{C} \left[~\overline{\nn}(x;\theta)\,\right]~,
\label{eq:ansatz_fbpinn}
\end{equation}
where
\begin{equation}
\overline{\nn}(x;\theta) = \sum^{n}_{i} w_{i}(x) \cdot \mathrm{unnorm}  \circ N\!N_{i} \circ \mathrm{norm}_{i}(x)~,
\end{equation}
and $\nn_{i}(x;\theta_{i})$ is a separate neural network placed in each subdomain $\Omega_{i}$, $w_i(x)$ is a smooth, differentiable window function that locally confines each network to its subdomain, $\mathcal{C}$ is a constraining operator which adds appropriate \say{hard} constraints to the ansatz such that it satisfies the boundary conditions (following the same procedure described in Section~\ref{sec:hard_pinns}), $\mathrm{norm}_{i}$ denotes separate normalisation of the input vector $x$ in each subdomain, $\mathrm{unnorm}$ denotes a common unnormalisation applied to each neural network output, and $\theta=\{\theta_{i}\}$. 


The purpose of the window function is to locally confine each neural network solution, $\nn_{i}$, to its subdomain. Any differentiable function can be used, as long as it is (negligibly close to) zero outside of the subdomain and greater than zero within it. For the hyperrectangular subdomains studied in this paper, we use the following window function;
\begin{equation}
w_i(x) = \prod^{d}_{j} \phi((x^{j} - a^{j}_{i})/\sigma^{j}_{i}) \phi((b^{j}_{i} - x^{j})/\sigma^{j}_{i})~,
\label{eq:window}
\end{equation}
where $j$ denotes each dimension of the input vector, $a^{j}_{i}$ and $b^{j}_{i}$ denote the midpoint of the left and right overlapping regions in each dimension (where $a^{j}_{i} < b^{j}_{i}$, i.e. the black lines in Figure~\ref{fig:workflow} (a)), $\phi(x) = \frac{1}{1+e^{-x}}$ is the sigmoid function, and $\sigma^{j}_{i}$ is a set of parameters defined such that the window function is (negligibly close to) zero outside of the overlap region. The individual normalisation $\mathrm{norm}_{i}$ in each subdomain is applied by normalising the input vector between $[-1,1]$ in each dimension over the subdomain before it is input to the network, whilst the common output unnormalisation $\mathrm{unnorm}$ is chosen such that each neural network output stays within the range $[-1,1]$, and depends on the solution itself. Any neural network can be used to define $\nn_{i}(x;\theta_{i})$; in this paper for simplicity we only consider fully connected neural networks. 

Given the ansatz defined by Equation~\ref{eq:ansatz_fbpinn}, FBPINNs are trained using the unconstrained loss function
\begin{equation}
\mathcal{L}(\theta) = \mathcal{L}_{p}(\theta) = \frac{1}{N_p}\sum^{N_p}_{i} ||\,\mathcal{D}[\hat u (x_i;\theta); \lambda] - f(x_i)\,||^2~,
\label{eq:fbpinn}
\end{equation}
using a set of training points $\{x_{i}\}$ sampled over the full domain $\Omega$. This loss function is the same form as used when training the strongly-constrained PINNs described in Section~\ref{sec:hard_pinns}, and doesn't require the use of additional interface terms by construction of our ansatz. Alternatively, the constraining operator $\mathcal{C}$ in the FBPINN ansatz can be removed and the corresponding \say{weak} loss function (cf Equation~\ref{eq:pinn}) used to train FBPINNs, in a similar fashion to PINNs.

\subsection{Flexible training schedules}
\label{sec:methods_schedules}

Alongside the issues with scaling PINNs to large domains described in Section~\ref{sec:motivation}, another issue is the difficulty of ensuring that the PINN solution learnt far away from the boundary is consistent with the boundary conditions. More precisely, it is conceivable that, early-on in training, the PINN could fixate on a particular solution which is inconsistent with the boundary condition at a location in the domain far away from the boundary, because the network has not yet learnt the consistent solution closer to the boundary to constrain it. With two different particular solutions being learned, the optimisation problem could end up in a local minima, resulting in a harder optimisation problem. Indeed, we find evidence for this effect in our numerical experiments (in particular, Sections~\ref{sec:1d_sin}~and~\ref{sec:wave}). Thus, for some problems it may make sense to sequentially learn the solution \say{outwards} from the boundary, in a similar fashion to, for example, time-marching schemes employed in classical methods. 

FBPINNs allow for such functionality, which is easy to implement because of their domain decomposition. At any point during training, we can restrict which subdomain networks are updated, and can therefore design flexible training schedules to suit a particular boundary problem. An example training schedule is shown in Figure~\ref{fig:active}. Within a training schedule we define \say{active} models, which are the networks which are currently being updated, \say{fixed} models, which are networks which have already been trained and have their free parameters fixed, and \say{inactive} models which are as-of-yet untrained networks. During each training step, only active models and their fixed neighbours contribute to the summation in the FBPINN ansatz, and only training points within the active subdomains are sampled.

\subsection{Parallel implementation}
\label{sec:parallel}

The FBPINN optimisation problem defined by Equation~\ref{eq:fbpinn} is solved by using gradient descent, similar to PINNs. This can be naively implemented within a single, global optimisation loop, but in practice, we can train FBPINNs in a highly parallel and more data-efficient way by taking advantage of the domain decomposition. In this section we describe a parallel implementation of FBPINNs, for which its pseudocode is shown in Figure~\ref{fig:threads}.

There are two key considerations when parallelising FBPINNs. First is that, outside of each neural network's subdomain its output is always zero after the window function has been applied, which means that training points outside of the subdomain will provide zero gradients when updating its free parameters. Thus, only training points within each subdomain are required to train each network, which allows training to be much more data-efficient. Second is that multiple parts of each training step can be implemented in parallel; a separate thread for each network can be used when calculating their outputs and gradients with respect to the input vector, and once the network outputs in overlapping regions have been summed, a separate thread for each network can be used to backpropagate the loss function and update their free parameters. This allows the training time to be dramatically reduced.

We now describe the parallel training algorithm in detail.  We use a standard gradient descent training algorithm, consisting of a number of identical gradient descent steps implemented inside a \verb+for+ loop. The pseudocode for each training step is shown in Figure~\ref{fig:threads}~(a), and a schematic of how the training step affects each subdomain is shown in Figure~\ref{fig:threads}~(b). Each training step consists of three distinct steps. First, for each subdomain, training points are sampled throughout the subdomain, normalised and input into the subdomain network. The output of the network is unnormalised, multiplied by the window function, and its appropriate gradients (depending on the specific problem) with respect to the input variables are computed using autodifferentiation. Second, for training points which intersect the overlapping regions between subdomains, the network outputs and gradients are shared across subdomains and summed. Third, for each subdomain, the constraining operator is applied to the summed solution and its gradients, the loss function is computed using these quantities, and the free parameters of the network are updated using backpropagation. We note that because each network has an independent set of free parameters, the computational graph can be discarded (or \say{detached}) for all outputs shared in its overlapping regions except for those from the current network when backpropagating, allowing the backpropagation operation to be implemented in parallel. Another note is that we must ensure that the same training data points are used in the overlapping regions for each network such that their solutions can be summed.


\section{Numerical experiments}
\label{sec:results}

\subsection{Overview of experiments}

In this section we carry out a number of experiments to test the accuracy and efficiency of FBPINNs. We are interested in how FBPINNs scale to larger problem sizes, and our experiments range from smaller to larger problems, both in terms of their domain size and dimensionality. Problems with multi-scale solutions are included.

First, in Section~\ref{sec:1d}, FBPINNs are tested on the motivating 1D example problem presented in Section~\ref{sec:motivation} (that of learning the solution to $\frac{du}{dx} = \cos \omega x$). Harder versions of this problem are introduced, including one with a multi-scale solution and another using second order derivatives in the underlying differential equation. In summary we find that FBPINNs are able to accurately and efficiently learn the solutions to these problems when $\omega$ is high, significantly outperforming PINNs. In Section~\ref{sec:2d}, we extend the motivating 1D problem to 2D, learning the solution to the equation $\frac{\partial u}{\partial x_1} + \frac{\partial u}{\partial x_2} = \cos(\omega x_1) + \cos(\omega x_2)$ when $\omega$ is high. Again, the FBPINN significantly outperforms the PINN tested. In Section~\ref{sec:burgers}, we test FBPINNs on a standard PINN benchmark problem, which is the (1+1)D viscous time-dependent Burgers equation. In this case the FBPINN matches the accuracy of the PINN tested, whilst being significantly more data-efficient. Finally in Section~\ref{sec:wave} we learn the solution to the (2+1)D time-dependent wave equation for a high-frequency point source propagating through a medium with a non-uniform wave speed, which is the most challenging problem studied here. Whilst the PINN tested exhibits unstable convergence, the FBPINN using a time-marching training schedule is able to robustly converge to the solution.

Some of the FBPINN settings and hyperparameters are fixed across all experiments. Specifically, the same optimiser (Adam \citep{Kingma2014}), learning rate (0.001), network type (fully connected), network activation function ($\tanh$), type of subdomain division (hyperrectangular), and window function (as defined in Equation~\ref{eq:window}) are used across all experiments. The relevant corresponding aspects are also fixed and the same across all of the PINN benchmarks used. Other settings and hyperparameters, such as the number of training steps, number of training points, number of hidden layers, number of hidden units, number of subdomains, overlapping width of each subdomain, output unnormalisation and training schedule vary depending on the problem, and for some cases studies we show ablations of them. For clearer comparison, we always use the same training point sampling scheme and density for both PINNs and FBPINNs, the same unnormalisation of their network outputs and the same constraining operator when forming their ansatze. The PINN input variable is always normalised between $[-1,1]$ in each dimension across the problem domain before being input to the network. The PINNs are implemented within the same coding framework as the FBPINNs, which uses the \verb+PyTorch+ library \citep{pytorch}. All 1D problems are trained using a single CPU core, whilst all other problems are trained using a single NVIDIA Titan V GPU. For this work we only use a single thread when training FBPINNs, although this thread does exactly implement the parallel training algorithm described in Section~\ref{sec:parallel}. Evaluating the multi-threaded performance of our parallel algorithm will be the subject of future work.


\subsection{1D sinusoidal experiments}
\label{sec:1d}

\begin{figure}[t]
\begin{center}
\includegraphics[width=15cm]{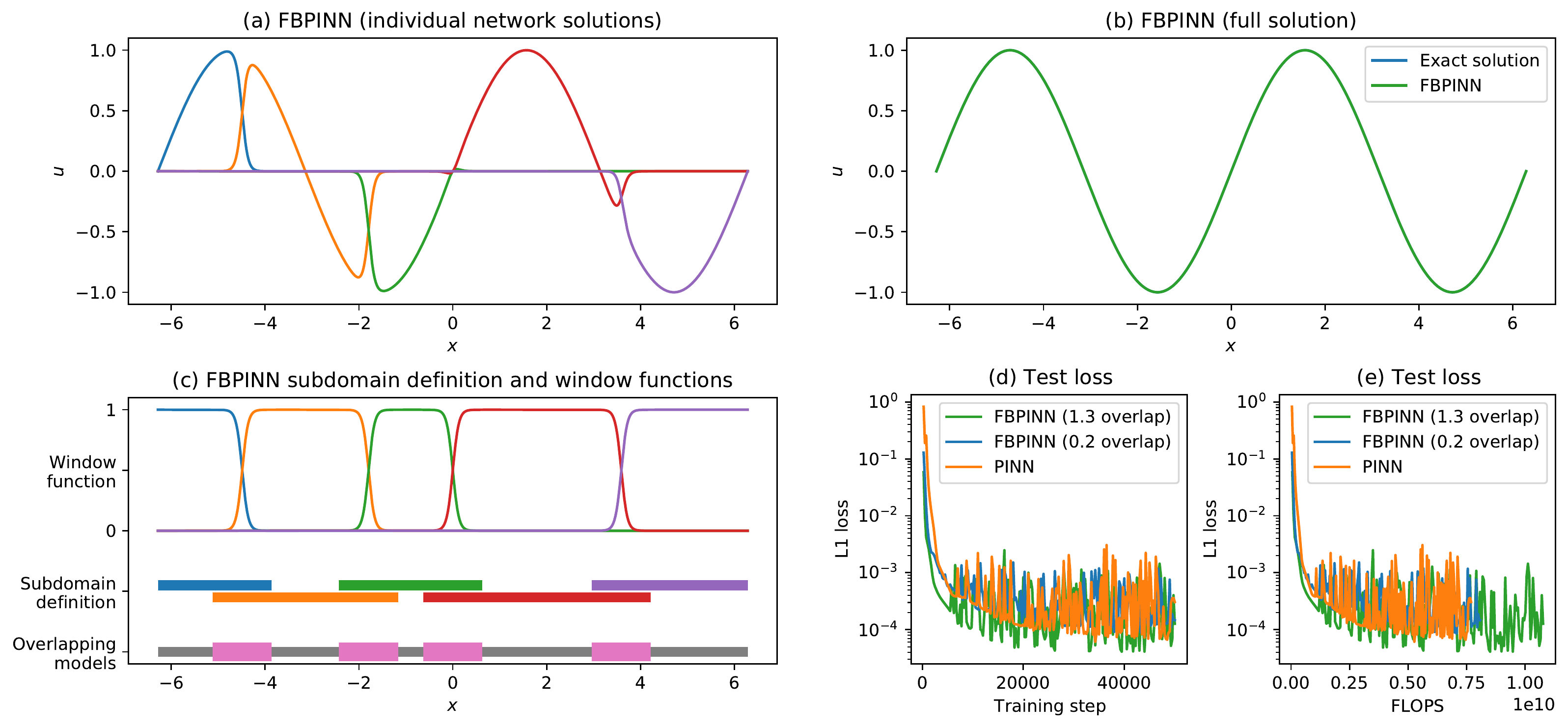}
\caption[]{Performance of FBPINNs on the motivating problem $\frac{d u}{d x} = \cos(\omega x)$ when $\omega=1$. For this case, we find the FPINN described in Section~\ref{sec:1d_low} has similar performance to the $\omega=1$ PINN in Section~\ref{sec:motivation_low} (shown in Figure~\ref{fig:cosp} (a)). The individual FBPINN subdomain solutions after training are shown in (a). The full FBPINN solution compared to the exact solution is shown in (b). The FBPINN subdomain definition, overlapping regions in the domain (thick pink lines), and window function for each subdomain are shown in (c). The L1 error between the FBPINN solution and the exact solution against training step is shown in (d) and (e). Also shown in (d) and (e) are the convergence curves for a FBPINN trained with a smaller subdomain overlap width and the $\omega=1$ PINN from Section~\ref{sec:motivation_low}.}
\label{fig:cos1f}
\end{center}
\end{figure}

\begin{figure}[t]
\begin{center}
\includegraphics[width=15cm]{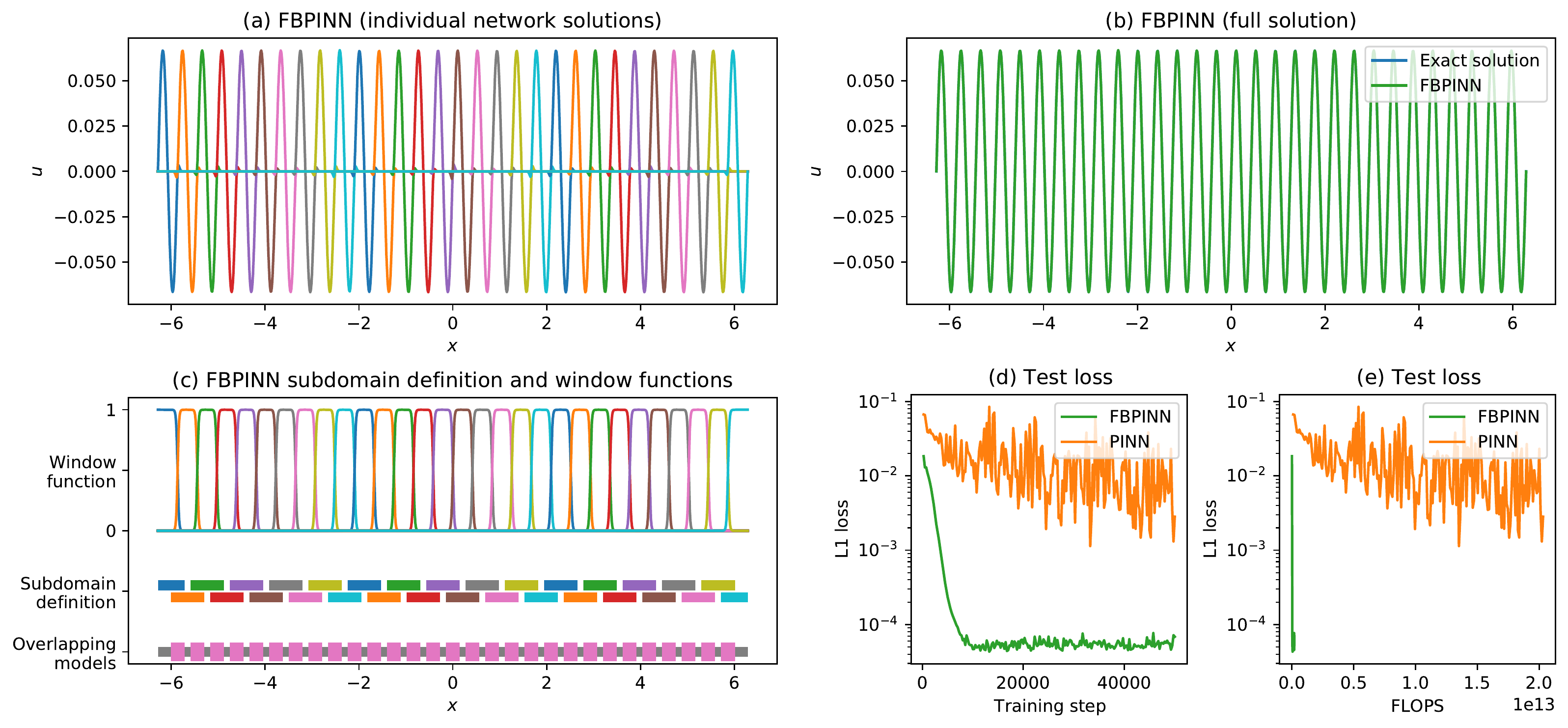}
\caption[]{Performance of FBPINNs on the motivating problem $\frac{d u}{d x} = \cos(\omega x)$ when $\omega=15$. The FBPINN described in Section~\ref{sec:1d_high} is compared to the best-performing $\omega=15$ PINN in Section~\ref{sec:motivation_high} (namely the PINN with 5 layers and 128 hidden units, shown in Figure~\ref{fig:cosp} (d)). For this case, we find the FPINN significantly outperforms the PINN, converging to the solution with much higher accuracy and much less training steps. This plot has the same layout as Figure~\ref{fig:cos1f}.}
\label{fig:cos15f}
\end{center}
\end{figure}

\begin{figure}[t]
\begin{center}
\includegraphics[width=15cm]{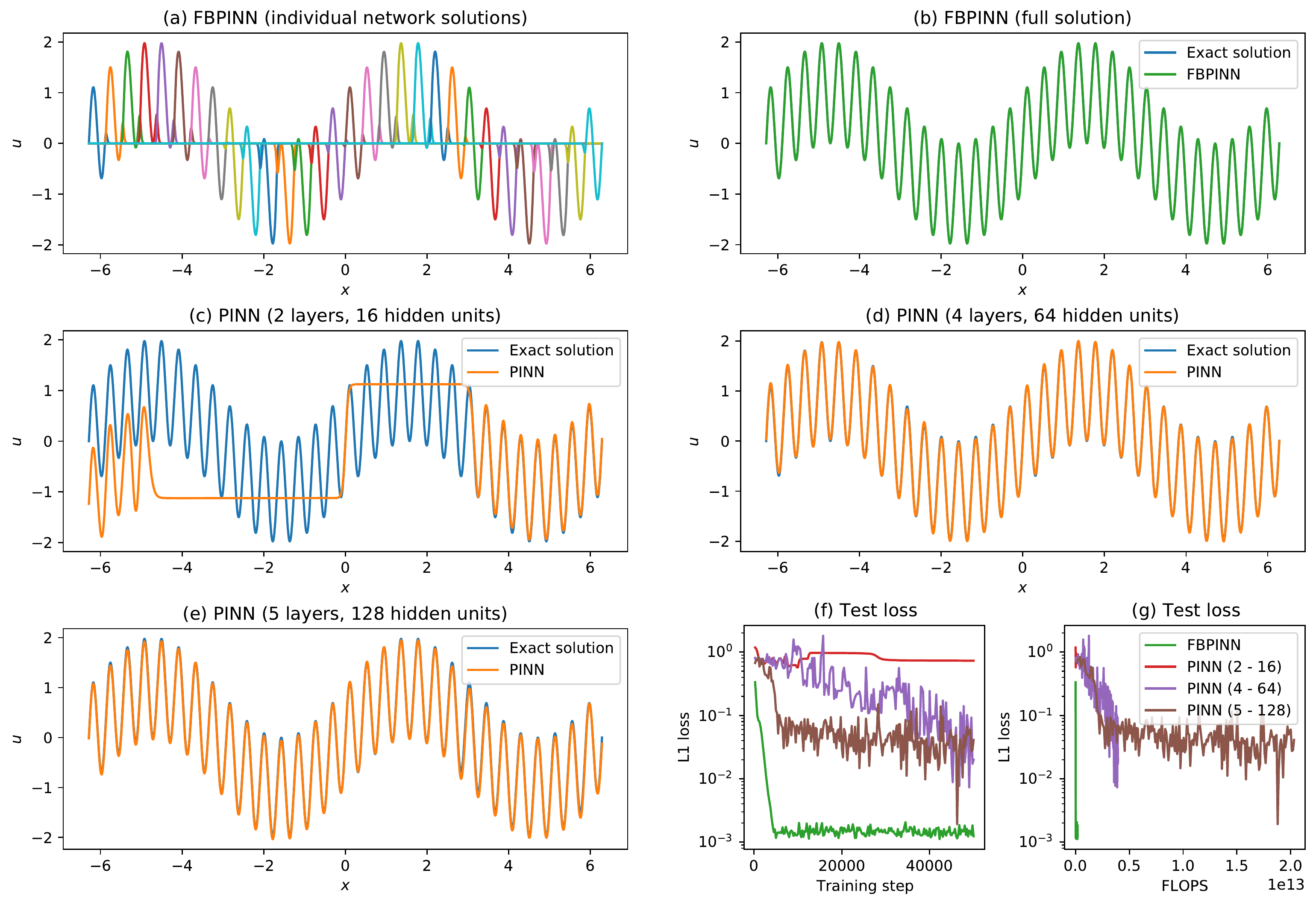}
\caption[]{Performance of FBPINNs on the multi-scale problem $\frac{du}{dx} = \omega_{1} \cos(\omega_{1} x) + \omega_{2} \cos(\omega_{2} x)$ where $\omega_1=1$ and $\omega_2=15$. The FBPINN described in Section~\ref{sec:1d_multi} is compared to three different PINNs which have 2 layers and 16 hidden units, 4 layers and 64 hidden units, and 5 layers and 128 hidden units. Similar to Figure~\ref{fig:cos15f}, the FBPINN significantly outperforms the PINNs tested. The individual FBPINN subdomain solutions after training are shown in (a). The full FBPINN solution is shown in (b). The three PINN solutions are shown in (c)-(e). The L1 errors of the FBPINN and PINN solutions compared to the exact solution are shown in (f) and (g).}
\label{fig:cosmulti}
\end{center}
\end{figure}

\begin{figure}[t]
\begin{center}
\includegraphics[width=15cm]{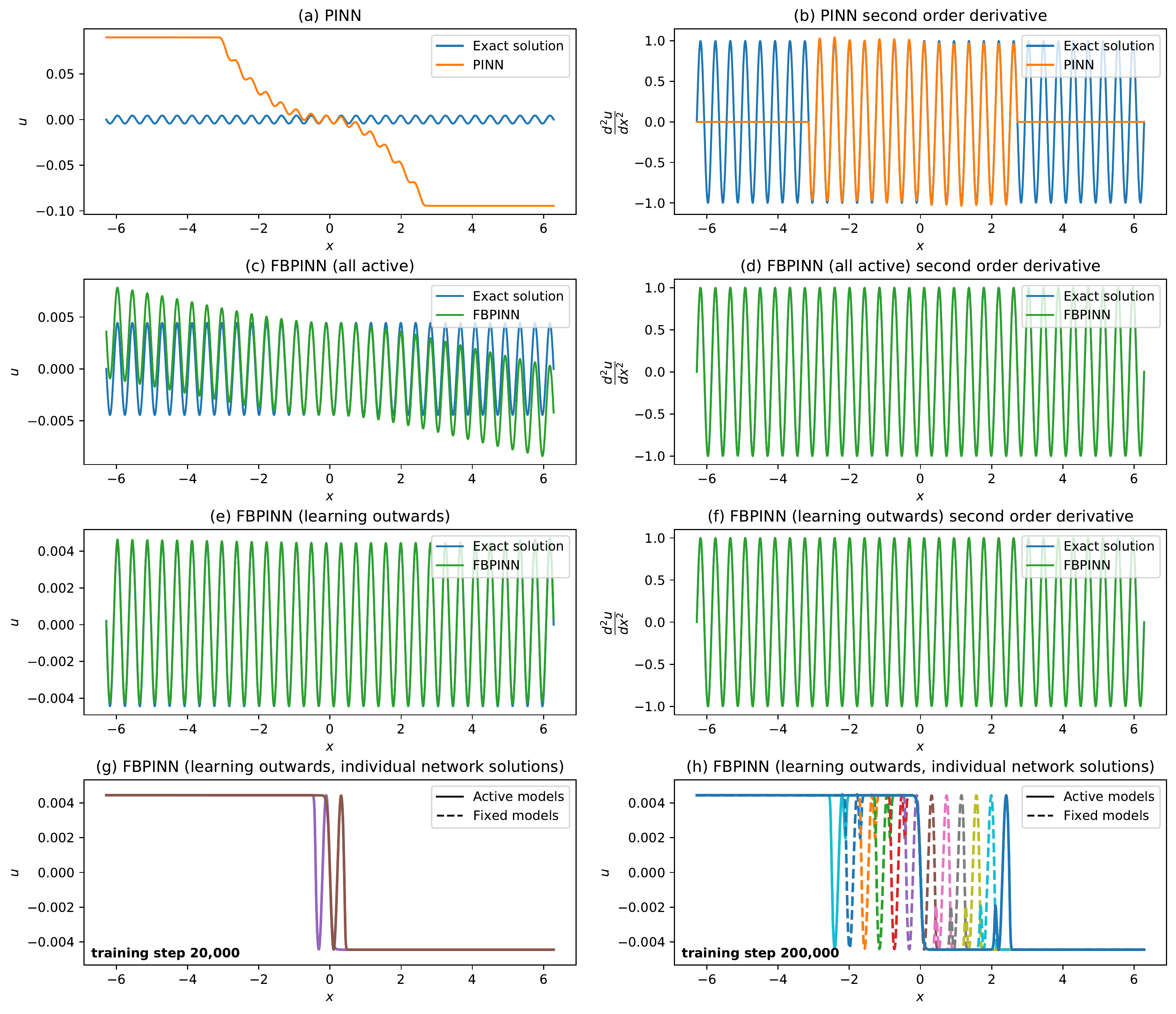}
\caption[]{Performance of FBPINNs on the problem $\frac{d^2u}{dx^2} = \sin(\omega x)$ with $\omega=15$. Two FBPINNs with different training schedules are tested for this problem. The first has an \say{all-active} training schedule, where all models are active all of the time, and its resulting solution and second order derivative are shown in (c) and (d). The second uses a \say{learning outwards} training schedule which slowly expands the active model outwards from the boundary condition at $x=0$ as training progresses (as depicted in (g) and (h)), and its solution and second order derivative are shown in (e) and (f). Both FBPINNs use the subdomain definition shown in Figure~\ref{fig:cos15f} (c), and are compared to a PINN with 5 layers and 128 hidden units, shown in (a) and (b).}
\label{fig:sin15}
\end{center}
\end{figure}

First, we test FBPINNs using the motivating 1D problem described in Section~\ref{sec:motivation} (Equation~\ref{eq:problem_motivate}). The following FBPINN ansatz is used
\begin{equation}
\hat u(x; \theta) = \tanh(\omega x) \overline{\nn}(x;\theta)~,
\end{equation}
which uses the same constraining operator as the PINN ansatz defined in Equation~\ref{eq:ansatz_motivate}. 

\subsubsection{Low frequency case ($\omega=1$)}
\label{sec:1d_low}

For the low frequency case, the domain $x\in [-2 \pi,2 \pi]$ is divided into $n=5$ overlapping subdomains shown in Figure~\ref{fig:cos1f}~(c). Each subdomain is defined such that all of their overlapping regions have a width of 1.3 and their associated window functions (as defined in Equation~\ref{eq:window}) are also shown in Figure~\ref{fig:cos1f}~(c). Each subdomain network has 2 hidden layers and 16 hidden units per layer. Similar to the PINN in Section~\ref{sec:motivation_low}, the output of each subdomain network is unnormalised by multiplying it by $\frac{1}{\omega}$ before summation, and the FBPINN is trained using 50,000 training steps and 200 training points regularly spaced over the domain. An \say{all-active} training schedule is used, where all of the subdomain networks are active every training step. The FBPINN has 1,605 free parameters in total.

Figure~\ref{fig:cos1f}~(a) shows the individual network solutions after training and (b) shows the full FBPINN solution. For ease of interpretation, each individual network plot in (a) shows the output of each subdomain network just before summation but with the constraining operator ($\tanh(\omega x) \cdot$) applied. Figure~\ref{fig:cos1f}~(d) compares the L1 convergence curve of the FBPINN to the L1 convergence curve of the low-frequency PINN (with 2 layers and 16 hidden units) studied in Section~\ref{sec:motivation_low}.  Figure~\ref{fig:cos1f}~(e) shows the same curve against the cumulative number of training FLOPS required during forward inference of the subdomain networks, which is a measure of  data-efficiency\footnote{Note, this measure only counts FLOPS spent during the forward inference of the networks, and does not count FLOPS spent during gradient computation, backpropagation or any other part of the training algorithm. See Appendix~\ref{sec:appendix_flops} for the exact formula used.}. For this case study, we find that the FBPINN is able to solve the problem as accurately and with a similar data-efficiency to the PINN. 

For this case study we also test the sensitivity of the FBPINN to different subdomain overlap widths. Figure~\ref{fig:cos1f}~(d) and (e) show the convergence curve for the same FBPINN but with its subdomains defined such that all overlapping regions have a width of 0.2. In this case the FBPINN has similar performance.

\subsubsection{High frequency case ($\omega=15$)}
\label{sec:1d_high}

Next we test the performance of the FBPINN when $\omega=15$ which is a much harder problem, as discussed in Section~\ref{sec:motivation}. For this case we divide the domain into $n=30$ equally spaced subdomains with overlapping widths of 0.3, as shown in Figure~\ref{fig:cos15f}~(c). The subdomain network size is kept the same as the case above at 2 layers and 16 hidden units per layer, and the same \say{all-active} training schedule is used. The FBPINN has 9,630 free parameters in total. Similar to the high-frequency PINNs tested in Section~\ref{sec:motivation_high}, the number of regularly spaced training points is increased to $200\times15=3000$. We compare the FBPINN to the best performing high-frequency PINN from Section~\ref{sec:motivation_high}, namely the PINN with 5 layers and 128 hidden units.

Figure~\ref{fig:cos15f} shows the same plots as Figure~\ref{fig:cos1f} for this case. We find that, in stark contrast to the PINN, the FBPINN is able to convergence to the solution with very high accuracy in very few training steps. Furthermore training the FBPINN requires multiple orders of magnitude less forward inference FLOPS than the PINN. This is because it uses much smaller network sizes in each of its subdomains, dramatically reducing the amount of computation required.

\subsubsection{Multi-scale case}
\label{sec:1d_multi}

We extend the difficulty of this problem by including multi-scale frequency components in its solution. Specifically, we consider the modified problem
\begin{equation}
\begin{split}
\frac{du}{dx} &= \omega_{1} \cos(\omega_{1} x) + \omega_{2} \cos(\omega_{2} x)~,\\
u(0) &= 0~,
\end{split}
\end{equation}
which has the exact solution
\begin{equation}
u(x) = \sin(\omega_{1} x) + \sin(\omega_{2} x)~.
\end{equation}
Here we chose $\omega_{1}=1$ and $\omega_{2}=15$, i.e. the solution contains both a high and low frequency component.

The same FBPINN and PINN from Section~\ref{sec:1d_high} are retrained for this problem, except that their loss functions are modified to use the differential equation above, unnormalisation is applied to both by multiplying their network outputs by 2, and $\omega_{2}$ is used in their ansatz constraining operator ($\tanh(\omega_{2} x) \cdot$). For this case we also train PINNs with smaller network sizes, namely 2 layers and 16 hidden units, and 4 layers and 64 hidden units.

The FBPINN individual network solutions, the FBPINN full solution and the three PINN solutions are shown in Figure~\ref{fig:cosmulti} (a), (b), (c), (d) and (e) respectively, and their L1 convergence curves are compared in (f) and (g). Similar results are observed to Section~\ref{sec:1d_high}: in stark contrast to the PINNs, the FBPINN is able to converge to the solution with a much higher accuracy in a much smaller number of training steps. Whilst the PINNs with 4 and 5 layers are able to model all of the cycles in the solution, their accuracy is nearly two orders of magnitude worse than the FBPINN, and their convergence curve is much more unstable.

\subsubsection{Second order derivative case}
\label{sec:1d_sin}

We also extend the difficulty of this problem by changing the underlying equation from a first order differential equation to a second order equation. Namely, we consider the related problem
\begin{equation}
\begin{split}
\frac{d^2u}{dx^2} &= \sin(\omega x)~,\\
u(0) &= 0~,\\
\frac{du}{dx}(0) &= -\frac{1}{\omega}~,
\end{split}
\end{equation}
which has the exact solution
\begin{equation}
u(x) = -\frac{1}{\omega^2} \sin(\omega x)~.
\end{equation}
We use the FBPINN ansatz
\begin{equation}
\hat u(x; \theta) = -\frac{1}{\omega^2} \tanh(\omega x) + \tanh^2(\omega x) \overline{\nn}(x;\theta)~,
\end{equation}
such that the boundary conditions are satisfied, and the same construction for the PINN ansatz. Alike Section~\ref{sec:1d_high} we consider the high frequency case $\omega=15$.

The same FBPINN and PINN from Section~\ref{sec:1d_high} are retrained for this problem, except for the changes to the problem definition above, and that the outputs of the FBPINN and PINN networks are unnormalised by multiplying them by $\frac{1}{\omega^2}$. We also train both networks for twice as long (100,000 steps).

The resulting FBPINN and PINN solutions are shown in Figure~\ref{fig:sin15}~(c) and (a), along with their second order derivatives in (d) and (b). We find that both methods struggle to accurately model the solution, although the FBPINN is able to capture all of its cycles. Whilst both models learn the solution accurately in the vicinity of the boundary condition, they learn it poorly outside of it. One explanation is that both models are suffering from integration errors; a small error in the second order derivative (which is  being penalised in the loss function) can lead to large errors in the solution. Indeed, both models learn much more accurate second order derivatives, as seen in Figure~\ref{fig:sin15}. Another explanation is that away from the boundary the FBPINN and PINN are fixating on a different (and incorrect) particular solution of the underlying equation. In particular, away from the boundary the FBPINN solution appears to be superimposed with a linear function of the input variable, which is a feasible solution under this differential equation (but is not consistent with the boundary conditions).

To improve the FBPINN solution further, we retrain the FBPINN using a training schedule that allows the solution to be \say{learned outwards} from the boundary condition. The schedule starts with only the two models in the center of the domain being active, and then slowly expands the active model outwards in the positive and negative directions, fixing the previously active models behind them, as shown in Figure~\ref{fig:sin15}~(g) and (h). In this case 500,000 training steps are used (equating to 33,333 training steps per active model). The resulting solution and its second order derivative are shown in Figure~\ref{fig:sin15}~(e) and (f). We find that this FBPINN performs best, accurately modelling the solution many cycles away from the boundary condition, although small errors remain at the edges of the domain.

\subsection{2D sinusoidal experiments}
\label{sec:2d}

\begin{figure}[t]
\begin{center}
\includegraphics[width=17.5cm]{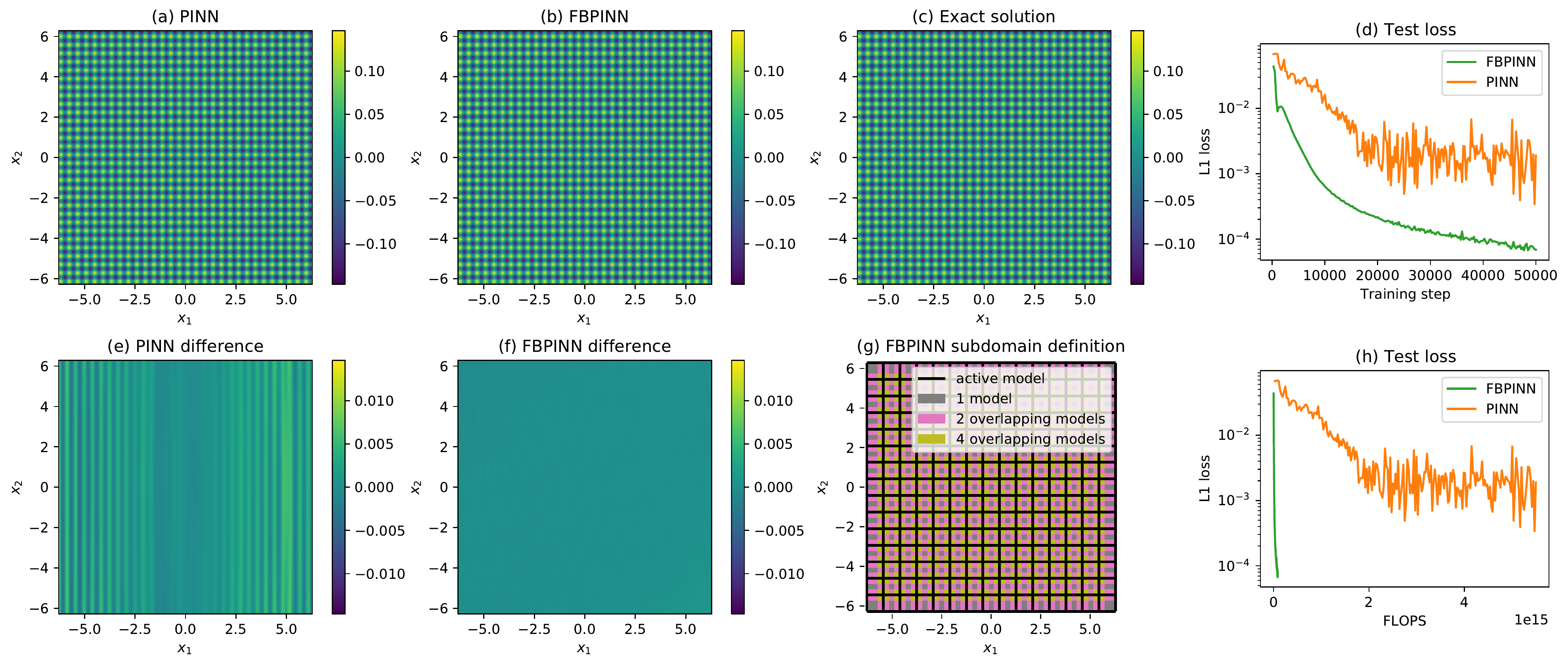}
\caption[]{Performance of FBPINNs on the problem $\frac{\partial u}{\partial x_1} + \frac{\partial u}{\partial x_2} = \cos(\omega x_1) + \cos(\omega x_2)$ with $\omega=15$. The FBPINN described in Section~\ref{sec:2d} is compared to a PINN with 5 layers and 128 hidden units. Similar to the 1D sinusodial problems above, we find that the FBPINN significantly outperforms the PINN tested. The FBPINN subdomain definition is shown in (g). The exact solution is shown in (c). The FBPINN solution and its difference to the exact solution are shown in (b) and (f), and a similar set of plots for the PINN are shown in (a) and (e). The L1 errors of the FBPINN and PINN solutions compared to the exact solution are shown in (d) and (h).}
\label{fig:cos_cos15}
\end{center}
\end{figure}

In this section we study the extension of the motivating problem above  from 1D to 2D. Specifically, we consider the problem
\begin{equation}
\begin{split}
\frac{\partial u}{\partial x_1} + \frac{\partial u}{\partial x_2} &= \cos(\omega x_1) + \cos(\omega x_2)~,\\
u(0,x_2) &= \frac{1}{\omega}\sin(\omega x_2)~,
\end{split}
\end{equation}
where $x = (x_1,x_2) \in \mathbb{R}^{2}, u \in \mathbb{R}^{1}$, with a problem domain $x_1 \in [-2 \pi,2 \pi],x_2 \in [-2 \pi,2 \pi]$. This problem has the exact solution
\begin{equation}
u(x_1,x_2) = \frac{1}{\omega} \sin(\omega x_1) + \frac{1}{\omega} \sin(\omega x_2)~.
\end{equation}
We use the FBPINN ansatz
\begin{equation}
\hat u(x_1, x_2; \theta) = \frac{1}{\omega} \sin(\omega x_2) + \tanh(\omega x_1) \overline{\nn}(x_1,x_2;\theta)~,
\end{equation}
such that the boundary conditions are satisfied, using the same constraining operator for the PINN ansatz. Similar to Section~\ref{sec:1d_high} we consider the high frequency case $\omega=15$. Note that the solution along the second dimension $x_2$ is already provided in the ansatz, and so the FBPINN and PINN only need to learn to correct the ansatz along the first dimension (although $x_2$ is still input to the networks and so they could still learn an incorrect solution along the second dimension). 

For the FBPINN we divide the 2D domain into $n = 15 \times 15 = 225$ equally spaced subdomains with overlapping widths of 0.6, as shown in Figure~\ref{fig:cos_cos15}~(g). Each subdomain network has 2 layers and 16 hidden layers, and we use the \say{all-active} training schedule defined above. For the PINN a network with 5 layers and 128 hidden units is chosen. The FBPINN has 75,825 free parameters whilst the PINN has 66,561 free parameters. The FBPINN and PINN are both unnormalised by multiplying their network outputs by $\frac{1}{\omega}$, and both are trained using $900\times900 = 810,\!000$ training points regularly spaced throughout the domain and 50,000 training steps. $1000\times1000$ regularly sampled test points throughout the domain are used when computing their L1 error compared to the exact solution.

Figure~\ref{fig:cos_cos15}~(c) shows the exact solution, (b) and (a) show the FBPINN and PINN solutions, and (f) and (e) show their difference to the exact solution. The FBPINN and PINN convergence curves are compared in Figure~\ref{fig:cos_cos15}~(d) and (h). Similar observations to Section~\ref{sec:1d_high} can be made, namely; the FBPINN is able to converge to the solution with much higher accuracy and much less training steps than the PINN; whilst the PINN is able to model all of the cycles of the solution, its accuracy is over one order of magnitude worse than the FBPINN; because a much smaller subdomain network size is used in the FBPINN, training the FBPINN requires multiple orders of magnitude less forward inference FLOPS than the PINN.

\subsection{(1+1)D Burgers equation}
\label{sec:burgers}

\begin{figure}[t]
\begin{center}
\includegraphics[width=17.5cm]{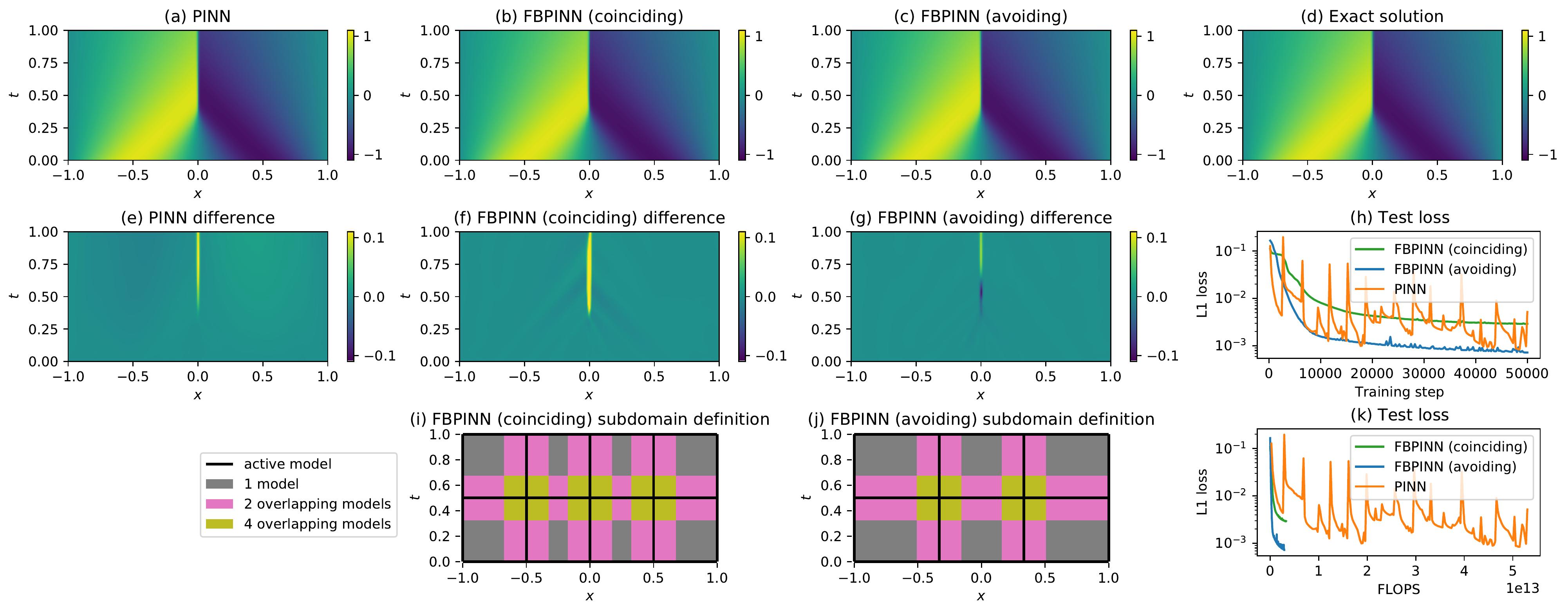}
\caption[]{Performance of FBPINNs on the (1+1)D viscous time-dependent Burgers equation. The exact solution is shown in (d). Two FBPINNs with different subdomain definitions are tested. The first uses a subdomain definition where the subdomain interfaces coincide with the discontinuity in the exact solution, shown in (i). The second uses a definition where the interfaces avoid the discontinuity, shown in (j). Both FBPINNs are compared to a PINN with 4 layers and 64 hidden units. The coinciding FBPINN solution and its difference to the exact solution are shown in (b) and (f). Similar sets of plots for the avoiding FBPINN and PINN are shown in (c) and (g), and (a) and (e) respectively. The L1 errors of the FBPINN and PINN solutions compared to the exact solution are shown in (h) and (k). We find that the coinciding FBPINN has slightly worse accuracy than the PINN, whilst the avoiding FBPINN has slightly better accuracy.}
\label{fig:burgers}
\end{center}
\end{figure}

In this section the FBPINN is tested using a standard PINN benchmark problem, which is the (1+1)D viscous time-dependent Burgers equation, given by
\begin{equation}
\begin{split}
\frac{\partial u}{\partial t} + u \frac{\partial u}{\partial x} &= \nu \frac{\partial^2 u}{\partial x^2}~,\\
u(x,0) &= - \sin(\pi x)~,\\
u(-1,t) &= 0~,\\
u(+1,t) &= 0~,
\end{split}
\end{equation}
where $x,t,u,\nu \in \mathbb{R}^{1}$, with a problem domain $x \in [-1,1],t \in [0,1]$. Interestingly, for small values of the viscosity parameter, $\nu$, the solution develops a discontinuity at $x=0$ as time increases. We use $\nu=0.01/\pi$ such that this is the case. The exact solution is analytically available by use of the Hopf-Cole transform (see \cite{Basdevant1986} for details, which are omitted here for brevity).

To solve the problem, we use the FBPINN ansatz
\begin{equation}
\hat u(x,t; \theta) = - \sin(\pi x) + \tanh(x+1)\tanh(x-1) \tanh(t)\overline{\nn}(x,t;\theta)~,
\end{equation}
such that the boundary conditions are satisfied, using the same constraining operator for the PINN ansatz.

For the FBPINN we divide the 2D domain into $n = 4 \times 2 = 8$ equally spaced subdomains with overlapping widths of 0.4, as shown in Figure~\ref{fig:burgers}~(i). We purposefully coincide the subdomain interfaces with the discontinuity in the solution at $x=0$, to test how the domain decomposition affects the solution accuracy across the discontinuity. Each subdomain network has 2 layers and 16 hidden layers, and we use the \say{all-active} training schedule. For the PINN a network with 4 layers and 64 hidden units is used, as in testing we found smaller networks performed worse. The FBPINN has 2,696 free parameters whilst the PINN has 12,737 free parameters. The FBPINN and PINN are both unnormalised by multiplying their network outputs by 1, and both are trained using $200\times200 = 40,\!000$ training points regularly spaced throughout the domain and 50,000 training steps. $400\times400$ regularly sampled test points throughout the domain are used when computing their L1 error compared to the exact solution.

Figure~\ref{fig:burgers} (d), (b) and (a) show the exact, FBPINN and PINN solutions respectively, and (f) and (e) show the difference of the FBPINN and PINN solutions to the exact solution. Figure~\ref{fig:burgers}~(h) and (k) show the L1 convergence curves of the FBPINN and PINN. We observe that the FBPINN solution is slightly less accurate across the discontinuity than the PINN, although its overall convergence is more stable. Whilst the FBPINN is able to model the discontinuity, it appears that the window function and summation of networks does make it harder for the FBPINN to model the solution in this region. To avoid this issue, we retrain the FBPINN using 6 overlapping subdomains with interfaces which do not coincide with the discontinuity, as shown in Figure~\ref{fig:burgers} (j). The resulting solution, difference and convergence curves are shown in (c), (g), (h) and (k), and we find that this FBPINN is able to more accurately model the discontinuity with a slightly higher overall accuracy than the PINN. Furthermore, as observed above, because a much smaller subdomain network size is used in the FBPINNs, training both FBPINNs requires multiple orders of magnitude less forward inference FLOPS than the PINN.

\subsection{(2+1)D Wave equation}
\label{sec:wave}

\begin{figure}[t]
\begin{center}
\includegraphics[width=13cm]{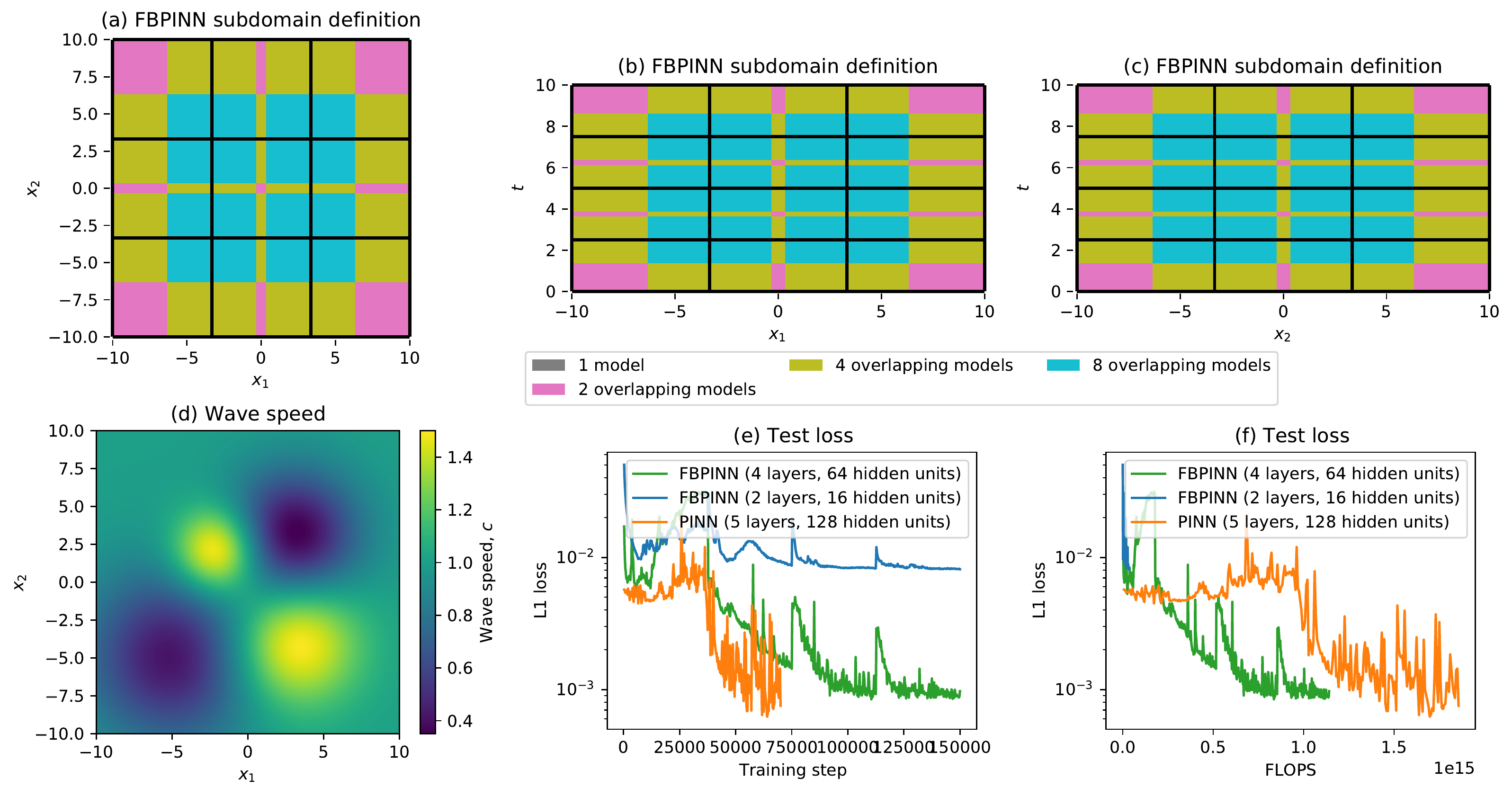}
\caption[]{Setup and convergence curves for the (2+1)D time-dependent wave equation problem. The FBPINN subdomain definition is shown in (a), (b), and (c), where the plots show orthogonal cross sections through the middle of the domain. The spatially-varying wave speed is shown in (d). The L1 errors of the FBPINN and PINN solutions compared to the solution from finite difference modelling are shown in (e) and (f).}
\label{fig:wave1}
\end{center}
\end{figure}

\begin{figure}[t]
\begin{center}
\includegraphics[width=17.2cm]{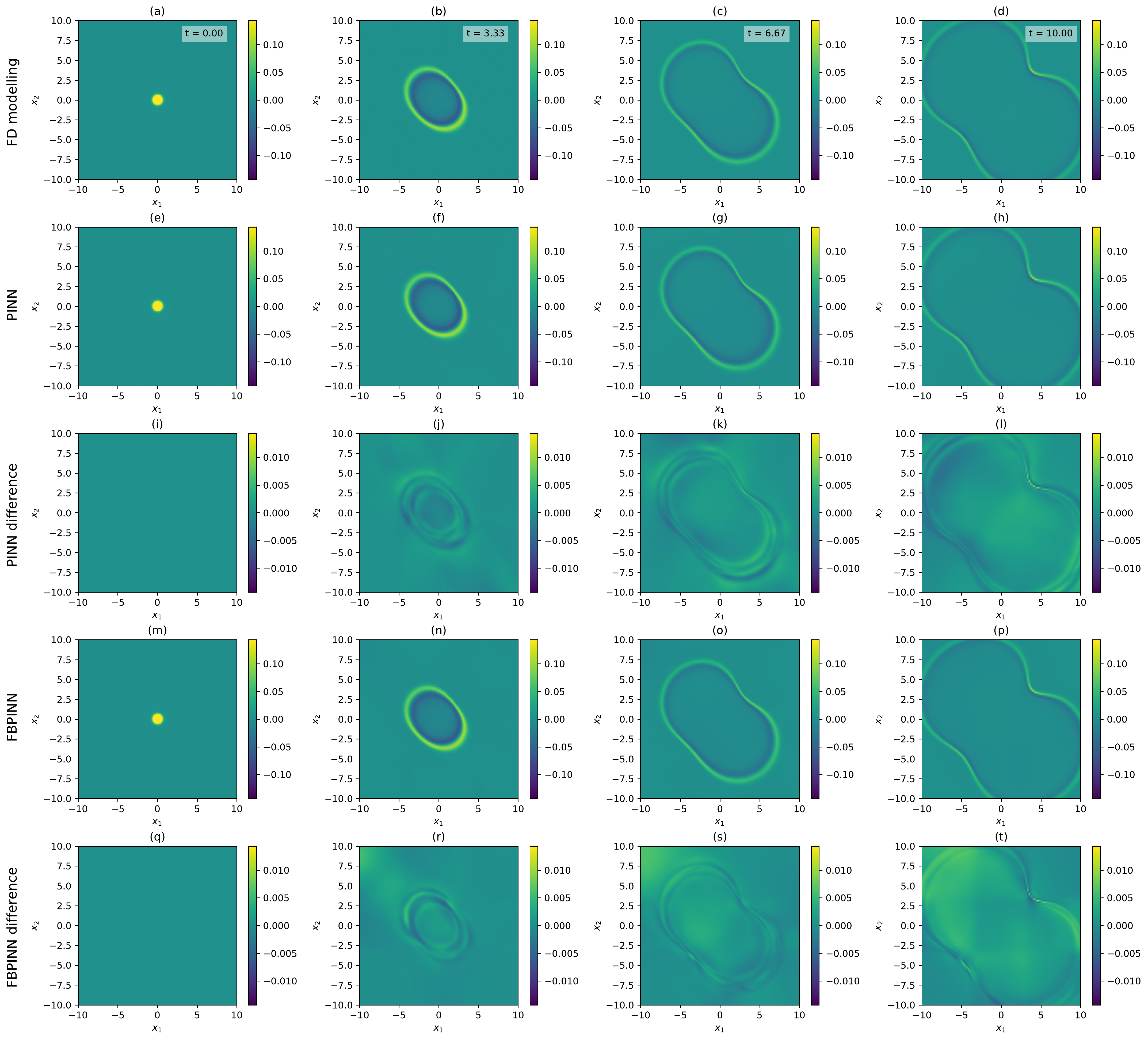}
\caption[]{Performance of FBPINNs on the (2+1)D time-dependent wave equation problem. The FBPINN described in Section~\ref{sec:wave} is compared to a PINN with 5 layers and 128 hidden units. We find that the FBPINN and PINN have similar accuracy, although the FBPINN converges more robustly to the solution (see Figure~\ref{fig:wave3}). The solution at 4 time-steps spanning the domain from finite difference modelling is shown in (a)-(d). The PINN solution at these time-steps is shown in (e)-(h), and the FBPINN solution at these time-steps is shown in (m)-(p). The difference of the PINN solution to the solution from finite difference modelling is shown in (i)-(l), and similar difference plots for the FBPINN are shown in (q)-(t).}
\label{fig:wave2}
\end{center}
\end{figure}

\begin{figure}[!ht]
\begin{center}
\includegraphics[width=17.2cm]{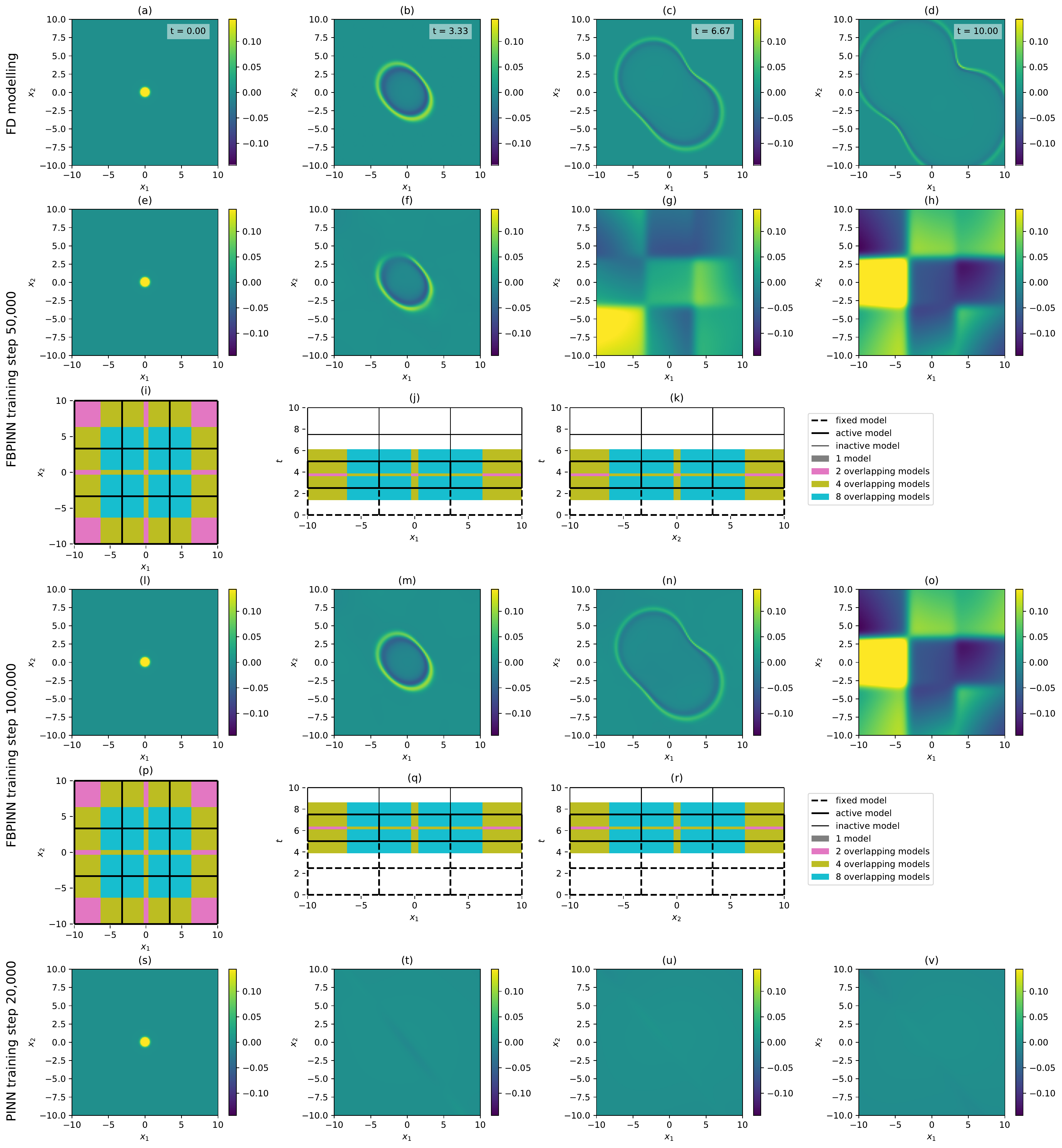}
\caption[]{FBPINN and PINN solutions during training for the (2+1)D time-dependent wave equation problem. The solution from finite difference modelling is shown in (a)-(d) (repeated from Figure~\ref{fig:wave2}). The FBPINN solution at training step 50,000 and 100,000 as well as the PINN solution at training step 20,000 are plotted in (e)-(h), (l)-(o) and (s)-(v) respectively using the same 4 time-steps. Whilst the FBPINN robustly learns the solution outwards from $t = 0$ as its time-marching training scheme progresses, even after 20,000 training steps the PINN solution is still close to zero everywhere apart from the boundary condition. Orthogonal cross sections of the active, fixed and inactive FBPINN subdomains through the center of the domain during the time-marching training schedule are shown for each FBPINN training step in (i)-(k) and (p)-(r).}
\label{fig:wave3}
\end{center}
\end{figure}

Finally, we test the FBPINN using the (2+1)D time-dependent wave equation, modelling the dynamics of an initially stationery point source propagating through a medium with a non-uniform wave speed. Specifically, we solve the following problem
\begin{equation}
\begin{split}
\left[ \nabla^{2} - \frac{1}{c(x)^2}\frac{\partial^2}{\partial t^2} \right] u(x,t) &= 0~,\\
u(x,0) &= e^{-\frac{1}{2}(||x-\mu||/\sigma)^2}~,\\
\frac{\partial u}{\partial t} (x,0) &= 0~,
\end{split}
\end{equation}
where $x,\mu \in \mathbb{R}^{2}$, $t,u,\sigma \in \mathbb{R}^{1}$, $c(x) \in \mathbb{R}^{1}$ is the spatially-varying wave speed, and $\mu$ and $\sigma$ control the starting location and central frequency of a Gaussian point source. The wave speed $c(x)$ is defined as a simple mixture of 2D Gaussian distributions and is shown in Figure~\ref{fig:wave1}~(d). For this case we use the problem domain $x_{1}\in[-10,10], x_{2}\in[-10,10], t\in[0,10]$ and set $\mu=(0,0)$ and $\sigma=0.3$. Modelling the wave equation can be challenging in general because its solutions are typically broadband, oscillatory and dispersive in nature, and can include reflections, refractions, wavefront compression and expansion through different velocity regions and a large range of amplitudes \citep{Igel2017}. For this case, the solution (or \say{wavefield}) expands outwards from the point source, compressing and expanding as it moves through regions with different wave speeds. We compare our results to the solution from finite difference (FD) modelling (See Appendix~\ref{sec:appendix_fd} for our detailed implementation), which is shown in Figure~\ref{fig:wave2}~(a)-(d) for 4 time-steps spanning the domain.

To solve the problem, we use the FBPINN ansatz
\begin{equation}
\hat u(x,t; \theta) = \phi(5(2-t/t_1))\, e^{-\frac{1}{2}(||x-\mu||/\sigma)^2} +  \mathrm{tanh}^2(t/t_1)\overline{\nn}(x,t;\theta)~,
\end{equation}
where $\phi(\cdot)$ is the sigmoid function and $t_1=\sigma/c(\mu)$. This ansatz is designed such that at $t=0$ the sigmoid function is (negligibly close to) 1, allowing the ansatz to match the boundary conditions. At times $t \gg 2\,t_1$, the sigmoid function is (negligibly close to) 0, removing the point source term from the ansatz. This is done so that the FBPINN does not need to learn to correct for this part of the ansatz once the point source has expanded away from its starting location. The same constraining operator is used for the PINN ansatz.

For the FBPINN we divide the 3D domain into $n = 3 \times 3 \times 4 = 36$ subdomains with overlapping widths of 6 in the spatial dimensions and 2 in the time dimension, as shown in Figure~\ref{fig:wave1}~(a)-(c). Each subdomain network has 4 layers and 64 hidden layers. For this problem we use a \say{time-marching} training schedule, where initially only subdomains in the first time-step are active, before slowly expanding the active time-step outwards and fixing the earlier time-step models. For the PINN a network with 5 layers and 128 hidden units is chosen. The FBPINN has 460,836 free parameters whilst the PINN has 66,689 free parameters. The FBPINN and PINN are both unnormalised by multiplying their network outputs by 1, and both are trained using $58\times58\times58 = 195,\!112$ training points randomly sampled throughout the domain. The PINN is trained for 75,000 training steps, whilst the FBPINN is trained for 150,000 steps (equating to 37,500 training steps per active model). For this case we chose random sampling over regular sampling because the training point density is relatively low compared to the frequency of the solution and in testing this allowed for better convergence of both the FBPINN and PINN. $100\times100\times10$ regularly sampled test points throughout the domain are used when computing their L1 error compared to the finite difference solution.

Figure~\ref{fig:wave2} (e)-(h) and (m)-(p) show the resulting PINN and FBPINN solutions respectively over 4 time-steps spanning the problem domain. Figure~\ref{fig:wave2} (i)-(l) and (q)-(t) show their difference compared to the finite difference solution, and Figure~\ref{fig:wave1}~(e) and (f) show their L1 convergence curves. We find that the FBPINN and PINN solutions have a similar accuracy, although the FBPINN takes roughly half as many forward inference FLOPS to train as the PINN, because its subdomain network size is smaller. We also test a FBPINN using a smaller subdomain network size (2 layers and 16 hidden units, the same as all previous examples) but this does not convergence, as shown in the convergence plots in Figure~\ref{fig:wave1}~(e) and (f). Whilst the PINN appears suitable for this problem, we plot the PINN and FBPINN solutions midway through training in Figure~\ref{fig:wave3}. We find that whilst the FBPINN robustly learns the solution outwards from $t=0$ as its time-marching training scheme progresses, even after 20,000 training steps the PINN solution is still close to zero everywhere apart from the boundary condition. This can also been seen in the convergence curve for the PINN in Figure~\ref{fig:wave1}~(e), where its L1 error actually increases until approximately training step 40,000. One explanation for this is that the PINN is fixating on a different (incorrect) particular solution away from the boundary early-on in training, namely the easier particular solution $u(x,t)=0$, causing the optimisation to become stuck in a local minima.


\section{Discussion}
\label{sec:discussion}

The numerical tests above confirm that FBPINNs provide a promising approach for scaling PINNs to large domains. They are able to accurately solve all of the smaller and larger scale problems studied, whilst in many cases the standard PINN struggles. For the problems studied with smaller domains, such as the Burgers equation and low-frequency sinusoidal problem, the FBPINN generally matches the PINN in performance. For the problems with larger domains, such as the wave equation and high-frequency sinusodial case studies, the FBPINN outperforms the PINN. The largest differences are seen in the high-frequency sinusoidal problems, where across all tests the FBPINN converges with much higher accuracy and much less training steps than the PINN. For the wave equation problem, the FBPINN more robustly converges to the solution. These findings demonstrate that the combined use of domain decomposition, separate subdomain normalisation and flexible training schedules helps to alleviate some of the major issues related to scaling PINNs to large domains.

FBPINNs also appear to be more data-efficient than standard PINNs. Across all experiments we find that FBPINNs are able to converge using smaller network sizes in their subdomains than the network size required by PINNs. The total number of forward inference FLOPS required during training of FBPINNs only depends on the subdomain network size, and not the number of subdomains (see Appendix~\ref{sec:appendix_flops} for proof), and thus the FBPINNs studied here require much less computation than the PINNs to train them. This is likely because of their \say{divide and conquer} strategy; each subdomain appears to present an easier optimisation problem which only requires a small number of free parameters to solve. Indeed, FBPINNs with more subdomains and smaller network sizes than those tested could be even more data-efficient, and in the future we plan to study in detail how reducing the subdomain size further affects their accuracy, and whether there is an optimal subdomain and network size to use (e.g. similar to h-p refinement in FEM).

It is important to note that each problem studied requires different configurations of the FBPINN to converge well. For example, for the high-frequency 1D first order sinusoidal problem a FBPINN with an \say{all-active} training schedule performs well, whilst for the second-order variant a FBPINN with a \say{learning outwards} training schedule is required to learn an accurate solution. Similarly, a subdomain network size of 2 layers and 16 hidden units is effective for every problem except the wave equation, which requires a larger network size given its subdomain definition. Thus it appears important to fine-tune the FBPINN to each particular problem. This may be because different scaling issues affect each problem differently. We also find that the FBPINN performs slightly worse for the Burgers equation problem when its subdomain interfaces coincide with the discontinuity in the solution, and therefore care must be taken when choosing the subdomain division.

Whilst we have focused on the issues related to scaling PINNs to large domains, another important consideration is the scaling of PINNs to higher dimensions. Similar to classical methods, a major challenge is likely to be the exponentially increasing number of (training) points required to sample the domain as the number of dimensions increases. It is important to note that domain decomposition may still help to reduce the complexity of the ensuing optimisation problem, and indeed FBPINNs are effective across all of the 1D, 2D and 3D problems studied. However, FBPINNs still require the same number of training points as standard PINNs and so issues such as the increased computational workload required are likely to remain. Specific to FBPINNs, the number of overlapping models summed in each overlapping region using the hyperrectangular subdivision grows exponentially with the number of dimensions, which could negatively affect the underlying FBPINN optimisation problem. We plan to investigate the scaling of FBPINNs to higher dimensions in future work.

A future direction is to study the performance of the multi-threaded version of FBPINNs in detail. For the single-threaded implementation of our parallel training algorithm used here, the FBPINNs are typically 2 to 10 times slower to train than their corresponding PINNs, despite the FBPINNs being more data-efficient. This is because the single thread updates each subdomain network sequentially, and also because each subdomain has a small network size and number of training points, meaning that the parallelism of the GPU is not fully utilised. The multi-threaded version (as described in Section~\ref{sec:parallel}) should reduce these training times by a factor proportional to the number of subdomains and yield a significant performance increase.

Another important direction is to test FBPINNs using irregular domains and subdomains. This is an essential step in many state-of-the-art classical approaches and FBPINNs are readily extendable in this regard. The same FBPINN framework can be used, and only the functions which sample points from each subdomain, define the neighbours of each subdomain and define the subdomain overlapping regions in the parallel training algorithm (Figure~\ref{fig:threads}~(a)) need to be changed. Going further, one could draw inspiration from classical methods where adaptive grids are used to solve multi-scale problems; it may be useful to adaptively change the subdomain definition and/or subdomain network in FBPINNs to dynamically fit to the solution. It is also important to note that in comparison to classical methods, where mesh refinement can be highly non-trivial, this could be relatively simple to implement using the mesh-free environment of FBPINNs.

Many other directions for applying and improving FBPINNs are possible. For example, FBPINNs could be used to solve inverse problems in the same way as PINNs and it would be interesting to compare their performance. There are many other types of differential equations which could be tested. Whilst we use simple fully connected subdomain networks here, other architectures and activation functions could be investigated. Another promising direction would be to use transfer learning within our flexible training schedules, for example by using neighbouring fixed models to initialise the free parameters of newly active models, which may improve accuracy and reduce training times further.

A major goal in the field of SciML is to provide ML tools which are practically useful for real-world problems and can extend or complement existing classical methods. For PINNs, one of the key remaining challenges is computational efficiency; training a PINN typically takes much more computational resources than using finite difference methods or FEM. Specifically for our wave equation problem in Section~\ref{sec:wave}, training the PINN / single-threaded FBPINN takes of the order of 10~hours on a single GPU, whilst FD modelling takes of the order of 1~minute on a single CPU. We noted above that the data-efficiency of FBPINNs increases as the size of its subdomain network decreases, and therefore with a small enough network size and a multi-threaded implementation FBPINNs may be able to match the efficiency of finite difference methods or FEM. It could also be powerful to combine efforts to learn families of solutions, such as the DeepONets mentioned above, with FBPINNs, allowing multiple large-scale solutions to be learnt without needing to retrain FBPINNs. Ultimately, this may lead to approaches that are faster and more accurate than classical methods, opening up many new potential applications. We also believe that standard benchmarks should be established to allow PINNs and their wide variety of derivatives to be more robustly compared, which will help the field achieve this goal.

\section{Conclusions}
\label{sec:conclusion}

In this work we presented FBPINNs, which are a scalable approach for solving large problems related to differential equations. By using a combination of domain decomposition, individual subdomain normalisation, and flexible training schedules, FBPINNs are able to alleviate some of the issues observed when scaling PINNs to large domains and/or multi-scale solutions, such as the increased complexity of the optimisation problem and the spectral bias of neural networks. We found that FBPINNs were able to accurately solve both the smaller and larger scale problems studied, including those with multi-scale solutions. Furthermore, FBPINNs are more data-efficient than PINNs, and they can be trained in a parallel fashion, which could eventually allow them to become more competitive with classical approaches such as finite difference or finite element methods. In future work we plan to study the performance of the multi-threaded version of FBPINNs, as well as adaptive subdomain refinement to further improve their accuracy and efficiency.

\paragraph{Acknowledgements}This work was funded by the UKRI EPSRC Center for Doctoral Training in Autonomous Intelligent Machines and Systems (AIMS CDT); we would like to thank the EPSRC and AIMS CDT for supporting this work. This work makes use of the BURGERS\_SOLUTION code written by John Burkardt (\href{https://people.sc.fsu.edu/~jburkardt/py_src/burgers_solution/burgers_solution.html}{link}) and the SEISMIC\_CPML library \citep{Komatitsch2007}.

\paragraph{Contributions (CRediT taxonomy)}BM: conceptualisation, formal analysis, investigation, methodology, software, validation, visualisation, writing - original draft.  AM, TNM: supervision, writing - review and editing.


\appendix

\section{Appendix}

\subsection{Forward inference FLOPS calculation}
\label{sec:appendix_flops}

To compare the computational resources required to train PINNs and FBPINNs, we use a measure which we call the forward inference FLOPS. This is a count of the cumulative number of FLOPS spent during forward inference of their neural network(s) during training (i.e., when computing the function $\nn(x;\theta)$ for PINNs or $\overline{\nn}(x;\theta)$ for FBPINNs), and therefore gives a measure of data-efficiency. 

For a fully connected neural network with $\tanh$ activations and a linear output layer, as used here, we assume the number of flops $F$ spent during forward inference of the network is
\begin{equation}
F = N ( (2d + 6)h + (l-1)(2h + 6)h + (2h + 1)d_u )~,
\label{eq:flops}
\end{equation}
where $N$ is the number of training points input to the network (or the batch size), $d$ is the dimensionality of the input vector, $d_u$ is the dimensionality of the output vector, $h$ is the number of hidden units and $l$ is the number of hidden layers. To see this, we note that the computation of each network layer consists of a matrix-matrix multiply, addition of the bias vector, and application of the activation function. For the first layer of the network, the matrix-matrix multiply requires $2Ndh$ operations, the bias addition requires $Nh$ operations and we assume the $\tanh$ operation requires $5Nh$ operations. Similar calculations follow for the remaining layers and hence Equation~\ref{eq:flops} follows. 

For PINNs, we use Equation~\ref{eq:flops} directly to estimate the forward inference FLOPS. For FBPINNs, the forward inference FLOPS are calculated using
\begin{equation}
F = \sum^n_i F_{i}~,
\label{eq:flops2}
\end{equation}
where the sum is over all subdomain networks used in the current training step.

We note that this measure only counts FLOPS spent during the forward inference of the networks, and does not count the additional FLOPS spent during gradient computation, backpropagation or any other part of the training algorithm.

\subsubsection{Scaling with network size / number of subdomains}

A noteable aspect of FBPINNs is that the total number of forward inference FLOPS required during training only depends on the subdomain network size, and not on the number of subdomains. We note that only training points within each subdomain are needed to train each subdomain network. We also note that, for a fixed domain and fixed training point density, as the number of subdomains increases the average number of training points falling over each subdomain decreases at the same rate. Therefore, using Equation~\ref{eq:flops2} and assuming the same network size per subdomain, it can be shown that the total number of forward FLOPS stays constant. However, this also assumes that the proportion of the domain covered by overlapping regions stays constant; the forward inference FLOPS will increase if this increases, as in these regions the outputs of all overlapping networks need to be computed.

\subsection{Finite difference modelling for (2+1)D Wave equation}
\label{sec:appendix_fd}

When studying the (2+1)D wave equation (Section~\ref{sec:wave}) we use finite difference modelling as the ground truth solution. The SEISMIC\_CPML library \citep{Komatitsch2007} is used to implement this (specifically, the seismic\_CPML\_2D\_pressure\_second\_order code), which performs staggered-grid second-order finite difference modelling of the time-dependent 2D acoustic wave equation using a convolutional perfectly matched layer (PML) boundary condition at the edges of the domain. For ease of use, we re-implemented the original Fortran code in Python. The simulation is initialised by sampling the spatially varying wave speed, initial wavefield and initial wavefield derivative as defined in Section~\ref{sec:wave} on a regular $694\times694\times1891$ grid ($x_1\times x_2\times t$). A high density of grid points ($\sim5\times$ spatial Nyquist frequency) is used so that the simulation is high-fidelity. An additional 10 grid points are used to define the PML boundary, which are cropped from the final solution before comparing it to the solution from the PINN and FBPINN.


\bibliographystyle{unsrtnat}
\bibliography{references}

\end{document}